\documentclass[a4paper,11pt]{article}
\usepackage{amsmath,mathtools,bm,amssymb,color}

\usepackage{graphicx}
\usepackage{subcaption}
\usepackage{jheppub} 
\usepackage{xspace}
\usepackage{booktabs}
\usepackage{array}
\usepackage{multirow}
\usepackage{adjustbox}

\renewcommand{\tabcolsep}{2em}

\newcommand{\pphllg}{\ensuremath{ H \rightarrow l^+l^-\gamma}\xspace}
\newcommand{\ppheeg}{\ensuremath{ H \rightarrow e^+e^-\gamma}\xspace}
\newcommand{\pphmumug}{\ensuremath{ H \rightarrow \mu^+\mu^-\gamma}\xspace}

\def\Q2{\left(Q^{2}\right)}

\def\l({\left(}
\def\r){\right)}

\def\gapprox{\lower .7ex\hbox{$\;\stackrel{\textstyle >}{\sim}\;$}}
\def\lapprox{\lower .7ex\hbox{$\;\stackrel{\textstyle <}{\sim}\;$}}

\newcommand{\nnlojet}{\texorpdfstring{NNLO\protect\scalebox{0.8}{JET}}{NNLOJET}\xspace}

\allowdisplaybreaks

\title{Fiducial cross sections for the lepton-pair-plus-photon decay mode in Higgs production up to NNLO QCD}

\author{X.\ Chen$^{a,b}$, T.\ Gehrmann$^{c}$, E.W.N.\ Glover$^d$, A.\ Huss$^e$}

\affiliation{
$^a$Institute for Theoretical Physics, Karlsruhe Institute of Technology, 76131 Karlsruhe, Germany\\
$^b$Institute for Astroparticle Physics, Karlsruhe Institute of Technology, 76344 Eggenstein-Leopoldshafen, Germany\\
$^c$Physik-Institut, Universit\"at Z\"urich, Winterthurerstrasse 190, CH-8057 Z\"urich, Switzerland\\
$^d$Institute for Particle Physics Phenomenology, Department of Physics, University of Durham, Durham, DH1 3LE, UK\\
$^e$Theoretical Physics Department, CERN, 1211 Geneva 23, Switzerland}

\emailAdd{xuan.chen@kit.edu}
\emailAdd{thomas.gehrmann@uzh.ch}
\emailAdd{e.w.n.glover@durham.ac.uk}
\emailAdd{alexander.huss@cern.ch}

\abstract{The rare three-body decay of a Higgs boson to a lepton-antilepton pair and a photon is starting to become 
experimentally accessible at the LHC. We investigate how higher-order QCD corrections to the dominant gluon-fusion 
production process impact on the fiducial cross sections in this specific Higgs decay mode for electrons and muons.
 Corrections up to NNLO QCD are 
found to be sizeable. They are generally uniform in kinematical variables related to the Higgs boson, but display several distinctive 
features in the kinematics of its individual decay products. 
}

\preprint{ZU-TH 52/21, KA-TP-23-2021, IPPP/21/42, P3H-21-078, CERN-TH-2021-165}

\begin{document}
\maketitle
\flushbottom

\section{Introduction}

The discovery of the Higgs boson~\cite{higgsexp1,higgsexp2} at the CERN LHC has initiated a large-scale research program that 
is aiming to determine the Higgs boson properties, 
its interactions with all Standard Model particles and with itself, as well as the dynamics of the Higgs sector of electroweak symmetry breaking in the 
Standard Model. This program proceeds mainly  through the study of multiple Higgs boson productions processes and decay modes, 
which are becoming increasingly accessible with the accumulation of the LHC data set. 

Among the different Higgs boson decay modes, the final state containing a lepton-antilepton pair and an identified photon, \pphllg,  has several  
peculiar features. Its tree-level contribution arises from the Yukawa coupling between the Higgs boson to the leptons. Owing to the 
smallness of the electron and muon Yukawa couplings, this contribution is highly suppressed, and the dominant processes leading to 
 \ppheeg and \pphmumug final states is the loop-mediated coupling of the Higgs boson to the lepton pair and the photon. The amplitudes 
 for this process involve the decay of the Higgs boson into 
 an electroweak gauge boson 
${\mathrm Z}/\gamma^*$ and a photon~\cite{Cahn:1978nz,Bergstrom:1985hp,Djouadi:1996yq}, as well as box 
contributions with a non-resonant coupling to the lepton pair~\cite{Abbasabadi:1996ze}. 
Provided the lepton mass can be neglected, the amplitude for this loop-induced process does not 
interfere with the Yukawa-induced tree-level amplitude. By selecting specific ranges in lepton pair invariant mass, the resonant ${\mathrm Z}/\gamma^*$ contribution
can be strongly enhanced, while the 
numerical effect of the box contributions becomes negligible. 
Given these restrictions, it is not sensible to define a branching fraction for $H\to Z\gamma$ or $H\to \gamma^*\gamma$ as a pseudo-observable in 
the Higgs sector. Instead, the fully differential three-body decay \pphllg should be studied over its phase space in 
the Higgs rest frame (Dalitz decay)~\cite{Passarino:2013nka}, applying dedicated cuts on invariant masses or particle energies to single out or 
enhance particular contributions of physics interest~\cite{Dicus:2013lta,Kachanovich:2020xyg,Kachanovich:2021pvx}. 

This is precisely what is being done in two recent ATLAS studies~\cite{ATLAS:2020qcv,ATLAS:2021wwb} that consider both high-mass (resonant-Z)
and low-mass ranges in the lepton pair invariant mass, with the resonant-Z region receiving an enhanced contribution from 
$H\to Z\gamma$ decays and the low-mass region being dominated by $H\to \gamma^*\gamma$ decays. 

Next-to-leading order (NLO) QCD corrections to the \pphllg decay mode affect only the top-quark loop contribution to the amplitude and are found 
to be moderate~\cite{Spira:1991tj,Gehrmann:2015dua,Bonciani:2015eua}. Higher-order QCD corrections are much more sizeable for the 
dominant Higgs boson production mode, gluon fusion. For the \pphllg decay mode, these have
up to now been considered in a 
fully differential manner up to NLO, as part of the MCFM code~\cite{Campbell:1999ah,Campbell:2011bn}. 
Fixed-order QCD corrections were computed up to 
N${}^3$LO~\cite{Anastasiou:2015vya,Mistlberger:2018etf,Dulat:2018bfe,Cieri:2018oms}
for Higgs boson production in gluon fusion 
and  
up to NNLO~\cite{Boughezal:2015dra,Boughezal:2015aha,Chen:2016zka,Chen:2018pzu,Campbell:2019gmd}
for Higgs-boson-plus-jet production.
These calculations are now being combined with the dominant decay modes ($\gamma \gamma$, $4l$, $2l2\nu$) to yield 
fully exclusive final states accurate to these 
orders~\cite{Anastasiou:2005qj,Catani:2007vq,Caola:2015wna,Bizon:2018foh,h4l,Chen:2021isd,Billis:2021ecs}. 

In the present paper, we investigate the impact of NNLO QCD corrections on fiducial cross sections related to \pphllg final states. Our predictions 
are obtained by including this decay mode in our earlier NNLO calculations for Higgs production and Higgs-plus-jet production in gluon fusion, 
as described in Section~\ref{sec:setup}. Numerical results are presented in Section~\ref{sec:results} for the  \pphllg decay mode, focusing on the fiducial regions 
defined by the ATLAS studies~\cite{ATLAS:2020qcv,ATLAS:2021wwb}. While the corrections 
are uniform in many of the kinematical variables, we observe perturbative instabilities 
in selected distributions. These could potentially be eased by modifications to the fiducial cuts, which we discuss in Section~\ref{sec:conc}.

\section{Setup}
\label{sec:setup}
The recent ATLAS studies~\cite{ATLAS:2020qcv,ATLAS:2021wwb} of the \pphllg decay at the 13 TeV LHC provided evidence for this 
decay mode in the range of low invariant masses of the di-lepton system. Our numerical investigation of higher-order 
QCD effects on the fiducial cross sections in this decay mode aims to reproduce the 
kinematical setup of the ATLAS physics analysis. 

Due to the scalar nature of the Higgs boson, its production and decay fully factorise. 
Using the narrow-width approximation for the Higgs propagator, we combine the dominant Higgs boson 
production process, gluon fusion, with the Higgs \pphllg  decay matrix elements. 
The gluon-fusion production matrix elements are calculated with the assumption of five massless quark flavours and an infinitely heavy top quark with effective $ggH$ coupling (HTL)~\cite{Wilczek:1977zn,Shifman:1978zn,Inami:1982xt}.
QCD corrections up to NNLO  are computed for Higgs production and Higgs-plus-jet production
using the \nnlojet framework, which is based on the 
antenna subtraction method~\cite{Gehrmann-DeRidder:2005btv,Daleo:2006xa,Currie:2013vh}, 
closely following our earlier work for other Higgs boson decay modes~\cite{Chen:2016zka,h4l}. 
 The \pphllg decay is described using leading-order matrix elements from loop-induced (closed quark and W boson loops) contribution of $({\mathrm Z}/\gamma^*)+\gamma$ production~\cite{Cahn:1978nz,Bergstrom:1985hp,Djouadi:1996yq}. The same-flavour-opposite-sign (SFOS) lepton pair is the  decay product from the ${\mathrm Z}/\gamma^*$. We only consider massless $e^\pm$ and $\mu^\pm$ in the final state throughout this paper. The implementation of \pphllg in \nnlojet is validated for Born-level predictions with up to three jets with MCFM~\cite{Campbell:2011bn,Campbell:1999ah}.
\begin{table}[t]
\centering
\begin{tabular}{@{\;} l <{\hspace{-1em}}>{\hspace{-1em}} c<{\hspace{-1em}} c @{\;}}
    \toprule
    Fiducial variables  & Low-mass region & Resonant-Z region\\
    \cmidrule(r){1-1}\cmidrule(lr){2-2}\cmidrule(l){3-3}
    $m_{l^+l^-}$ (GeV)   & [0.21, 30]  & [81.2, 101.2] \\ 
    $m_{e^+e^-}$ veto (GeV)   & [2.5, 3.5], [8, 11]& - \\ 
    $m_{\mu^+\mu^-}$ veto (GeV)   & [2.9, 3.3], [9.1, 10.6]& - \\ 
    $m_{l^+l^-\gamma}$ (GeV)  & 125 & 125 \\
    $p^{\gamma}_T$ (GeV)  & $>37.5$ & $>15$ \\
    $|y^{\gamma}|$  & $< 2.37$ & $< 2.37$ \\
    $p^{l^+l^-}_T$ (GeV)  & $>37.5$ & - \\
    1st lepton $p^{e_1(\mu_1)}_T$ (GeV)  & $>13$ (11) & $>10$ \\
    2nd lepton $p^{e_2(\mu_2)}_T$ (GeV)  & $>4.5$ (3) & $>10$ \\
    $|\eta^{e(\mu)}|$  & $< 2.47 (2.7)$ & $< 2.47 (2.7)$ \\
    $\Delta R(l^\pm,\gamma)$ & $> 0.4$ & $> 0.3$ \\
    \bottomrule
  \end{tabular}
   \caption{Fiducial selection criteria of \pphllg final states inspired by ATLAS analysis for near on-shell Z boson mass region~\cite{ATLAS:2020qcv} and low-mass region~\cite{ATLAS:2021wwb}.
   \label{tab:fid-cut}}
\end{table}

The ATLAS studies~\cite{ATLAS:2020qcv,ATLAS:2021wwb} consider fiducial cross sections for SFOS lepton pairs 
in two lepton-pair invariant mass ranges, named `low-mass' and `resonant-Z'. The fiducial selection criteria are summarised in 
Table~\ref{tab:fid-cut}. To avoid the divergent on-shell photon pole in the low-mass region, a minimum 
pair invariant mass of $m_{l^+l^-}$ of 210~MeV is required. In addition, $m_{l^+l^-}$ veto regions to eliminate background contributions of decay products from $J/\psi$ and $\Upsilon$ mesons are defined. Owing to the different resolution in the reconstruction of electrons and muons, the size of these veto
regions depends on the lepton flavour. 
To study the impact of fiducial selection criteria in Table~\ref{tab:fid-cut}, we define acceptance factors as the ratio of 
the fiducial cross sections divided by the cross sections for inclusive \pphllg production (identical for $e^+e^-$ and $\mu^+\mu^-$ channels) 
in the low-mass or resonant-Z region. These inclusive  \pphllg production cross sections are 
obtained by only applying the 
$m_{l^+l^-}$ cuts relevant to each mass region (first row of Table~\ref{tab:fid-cut}), i.e.\ without any further cuts on the lepton and photon 
momenta and no 
invariant mass vetoes. 

The predictions are obtained using the Higgs boson parameters:
mass $m_H=125$~GeV, decay width $\Gamma_H=4.088\times 10^{-3}$~GeV~\cite{LHCHiggsCrossSectionWorkingGroup:2016ypw} and vacuum expectation value $v = 246.2$~GeV.
The top quark mass in the pole scheme is set to $m_t=173.2$~GeV. We employ the $G_\mu$ scheme for electroweak couplings with $m_{\mathrm Z}= 91.19$~GeV, $\Gamma_{\mathrm Z}=2.495$~GeV and $m_{\mathrm W}= 80.38$~GeV. We use the NNPDF3.1 parton distribution functions (PDFs) at NNLO accuracy with the value of $\alpha_s(m_{\mathrm Z})$ = 0.118~\cite{NNPDF:2017mvq}. Note that we systematically use the same set of PDFs and the same value of $\alpha_s(m_{\mathrm Z})$ for LO, NLO and NNLO predictions. The central factorisation ($\mu_F$) and renormalisation ($\mu_R$) scales are fixed to be $m_H/2$. The theoretical uncertainty is estimated by varying $\mu_F$ and $\mu_R$ independently by a factor of two while ignoring the extreme combinations of $\mu_F/\mu_R$ being 1/4 or 4. This is the so-called 7-point scale variation and we present scale uncertainties by taking the envelope of the seven combinations.

Finite top quark mass effects in gluon fusion Higgs production are mandatory to obtain reliable predictions for precision 
phenomenology. These are included by  reweighting higher order corrections obtained in the HTL approximation by the 
 full top mass dependence at leading order. This procedure has 
 proven to be reliable at NLO~\cite{Harlander:2012hf,Neumann:2014nha,Neumann:2016dny,Becker:2020rjp,Chen:2021azt}, 
 where the  full top mass dependence of the inclusive Higgs production cross section~\cite{Spira:1995rr,Anastasiou:2009kn,Czakon:2021yub} 
 and its transverse momentum distribution~\cite{Jones:2018hbb,Lindert:2018iug,Neumann:2018bsx} are known. 
 In this paper, we adopt the multiplicative reweighting strategy at histogram level using LO fiducial total and differential cross sections with finite top mass corrections. We label the reweighed HTL results as HTL$\otimes$SM following the procedure described in~\cite{Chen:2016zka}.
The scales are chosen identically between the reweithging factor and the HTL predictions, i.e.\ in a fully correlated manner, in order to ensure that HTL$\otimes$SM results at LO exactly reproduce the LO predictions with full top mass dependence.

\section{Results}
\label{sec:results}
In this section, we present predictions for fiducial cross sections in the \pphllg decay channel at the LHC. We include up to NNLO QCD corrections for the gluon-fusion Higgs production channel and correct multiplicatively for finite top mass effects at LO. Applying the fiducial selection criteria listed in Table~\ref{tab:fid-cut},  we compute the fiducial total and differential cross sections for both low-mass and resonant-Z boson mass windows to determine the signal yield and shape of distributions for the \pphllg decay channel with final state electrons or muons.
We aim to provide theoretical benchmarks to increase the sensitivity of the LHC data analysis and to prepare  for the discovery of \pphllg decay mode.
\label{sec:result}
\subsection{Fiducial total cross sections}
\label{sec:fid-to-inc}
The total cross sections with fiducial cuts are documented in Table~\ref{tab:totalXS} for low-mass and resonant-Z regions with up to NNLO QCD corrections in the 
HTL approximation. Adopting the reweighting procedure introduced above, finite top quark mass effects are accounted for to LO accuracy for the  HTL$\otimes$SM predictions. We observe positive corrections of about 6.6\% after including top quark mass effects for both low-mass and resonant-Z regions. The relative size of scale uncertainties remains the same between HTL and HTL$\otimes$SM predictions. 

\begin{table}[t]
\centering
\begin{adjustbox}{width=1\textwidth}
\begin{tabular}{@{\;} l <{\hspace{-1em}}>{\hspace{-1em}} c@{\hspace{1em}}c@{\hspace{1em}}c <{\hspace{-1em}}>{\hspace{-1em}} c@{\hspace{1em}}c@{\hspace{1em}}c @{\;}}
    \toprule
    \multirow{2}{*}{\large$\qquad\quad\sigma$[fb]} & \multicolumn{3}{c}{Low-mass region} & \multicolumn{3}{c}{Resonant-Z region}\\
                 & HTL & HTL$\otimes$SM & Acceptance & HTL & HTL$\otimes$SM & Acceptance\\
    \cmidrule(r){1-1}\cmidrule(lr){2-4}\cmidrule(l){5-7}
    \ppheeg   @ LO   & $0.243^{+0.065}_{-0.046}$ & $0.259^{+0.069}_{-0.049}$ & $41.7\%$ & $0.425^{+0.114}_{-0.081}$  & $0.453^{+0.121}_{-0.086}$ & $56.3\%$\\
    \ppheeg   @ NLO  & $0.504^{+0.103}_{-0.079}$ & $0.537^{+0.109}_{-0.084}$ & $37.7\%$ & $0.976^{+0.221}_{-0.165}$  & $1.040^{+0.235}_{-0.175}$ & $56.3\%$\\
    \ppheeg   @ NNLO & $0.636^{+0.059}_{-0.063}$ & $0.678^{+0.063}_{-0.068}$ & $37.7\%$ & $1.229^{+0.110}_{-0.124}$ & $1.310^{+0.117}_{-0.133}$ & $56.3\%$\\
    \cmidrule(r){1-1}\cmidrule(lr){2-4}\cmidrule(l){5-7}
    \pphmumug @ LO   & $0.299^{+0.080}_{-0.057}$ & $0.319^{+0.085}_{-0.061}$ & $51.3\%$ & $0.451^{+0.121}_{-0.086}$  & $0.481^{+0.129}_{-0.091}$ & $59.8\%$\\
    \pphmumug @ NLO  & $0.613^{+0.124}_{-0.096}$ & $0.653^{+0.132}_{-0.102}$ & $45.8\%$ & $1.034^{+0.233}_{-0.174}$  & $1.102^{+0.249}_{-0.186}$ & $59.7\%$\\
    \pphmumug @ NNLO & $0.774^{+0.072}_{-0.077}$ & $0.825^{+0.077}_{-0.082}$ & $45.9\%$ & $1.302^{+0.116}_{-0.132}$ & $1.387^{+0.124}_{-0.140}$ & $59.7\%$\\
    \bottomrule
  \end{tabular}
\end{adjustbox}
   \caption{Fiducial total cross sections for \ppheeg and \pphmumug productions at the LHC in the HTL reweighted by SM predictions with LO top quark mass effect. Predictions include up to NNLO QCD correction from hadronic initial states. The upper and lower values correspond to the envelope of 7-point scale variations.~\label{tab:totalXS}}
\end{table}

For the \ppheeg channel, higher order QCD corrections to the gluon fusion process result in an enhancement of the signal yield in the low-mass region by  107\% at NLO and a further 26\% at NNLO. The scale uncertainty band decreases slightly from LO to NLO
and is reduced substantially to $\pm 10\%$ at NNLO. 
For the resonant-Z boson mass region, we observe a larger enhancement of the signal yield at NLO (by 130\%) and a
further 26\% enhancement at NNLO. The NNLO scale uncertainty amounts to $\pm 10\%$ as in the low-mass region. The signal yield in the resonant-Z region is about twice as large as in the low-mass region. 
The acceptance factor for the low-mass region decreases from $41.7\%$ at LO to $37.7\%$ at NLO and stabilizes at NNLO. This change 
in acceptance factor can be attributed to the sensitivity of the lepton and photon transverse momentum distributions on higher order QCD 
radiation, which will be discussed in more detail below. 
In the
 resonant-Z region, the acceptance factors are independent of higher order QCD corrections, being $56.3\%$ throughout at LO, NLO and NNLO.

For the \pphmumug channel, the fiducial selection criteria in Table~\ref{tab:fid-cut} yield a more inclusive coverage of the final state phase space. Reducing the mass veto windows in $m_{l^+l^-}$ and accepting muon pseudo-rapidities up to $\pm 2.7$
results in larger fiducial cross sections. The enhancement of the fiducial cross sections from electrons
to muons is more prominent for the low-mass region, where it amounts (at NNLO HTL$\otimes$SM accuracy) to $+22\%$, compared to
$+6\%$ in the resonant-Z region. The relative size of the higher-order QCD corrections and the associated
uncertainties are very similar to the \ppheeg channel.
The acceptance factors of  the \pphmumug channel are larger than for the \ppheeg channel as expected, with the 
enhancement being more prominent for the low-mass than in the resonant-Z region. We observe again that the acceptance factors 
depend on the perturbative order only in the low-mass region. 

The study of above fiducial total cross sections indicates that the signal yield between low-mass and resonant-Z regions are of a comparable order of magnitude, and they receive large and positive QCD corrections.
The acceptance factors behave very differently between low-mass and resonant-Z regions with
sensitivity to the higher order QCD corrections only in the low-mass region.

The dominant irreducible background for both mass regions is the ordinary continuum  Drell-Yan production with extra photon radiation.
For the resonant-Z region, the signal extraction is particularly challenging due to the expected low values
of photon transverse momentum. For Born-level kinematics and an on-shell Z-boson, the maximum photon transverse
momentum amounts to 29.2~GeV, with higher values attained
only if the Z-boson or the Higgs boson is off its mass shell. Consequently, the ATLAS search in~\cite{ATLAS:2020qcv}
incorporated an improved photon
identification efficiency in the $p_T^\gamma$ range of $10$ to $35$~GeV.
For the low-mass region, on the contrary, a dedicated trigger is applied in~\cite{ATLAS:2021wwb} for $p_T^\gamma>35$~GeV to reduce the continuum background. Without much loss of signal yield in the low-mass region, it appears that an
adequately large $p_T^\gamma$ cut is the key fiducial selection criterion towards the discovery of \pphllg channel. More details regarding the impact of fiducial selection criteria will be discussed by examining fiducial differential cross sections in the following.

\subsection{Fiducial differential cross sections}
\label{sec:fid-diff}

Measurements of differential cross sections for the \pphllg decay channel at the LHC are currently not available due to limited luminosity and large background contributions. Nevertheless, theoretical predictions for fiducial differential cross sections for this channel reveal kinematic information that could help the experimental analysis to define its signal regions and to improve signal-to-background ratios. Based on the fiducial cuts listed in Table~\ref{tab:fid-cut}, we present some selected differential observables in this section and discuss the impact of fiducial cuts on the size and shape of the distributions. 

\begin{figure}[t]
    \centering
    \begin{subfigure}[b]{0.49\textwidth}
        \includegraphics[width=\textwidth]{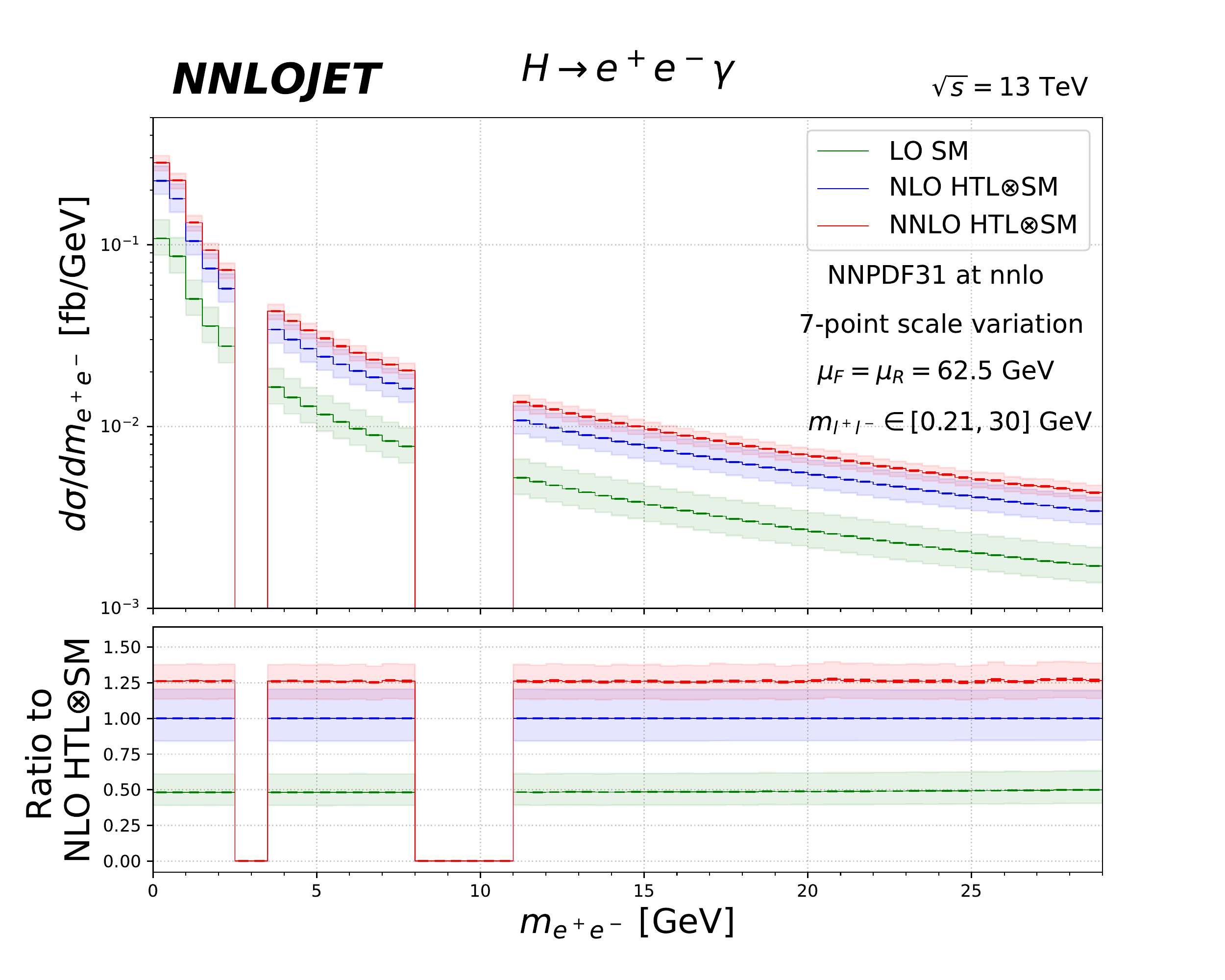}
    \end{subfigure}
    \begin{subfigure}[b]{0.49\textwidth}
        \includegraphics[width=\textwidth]{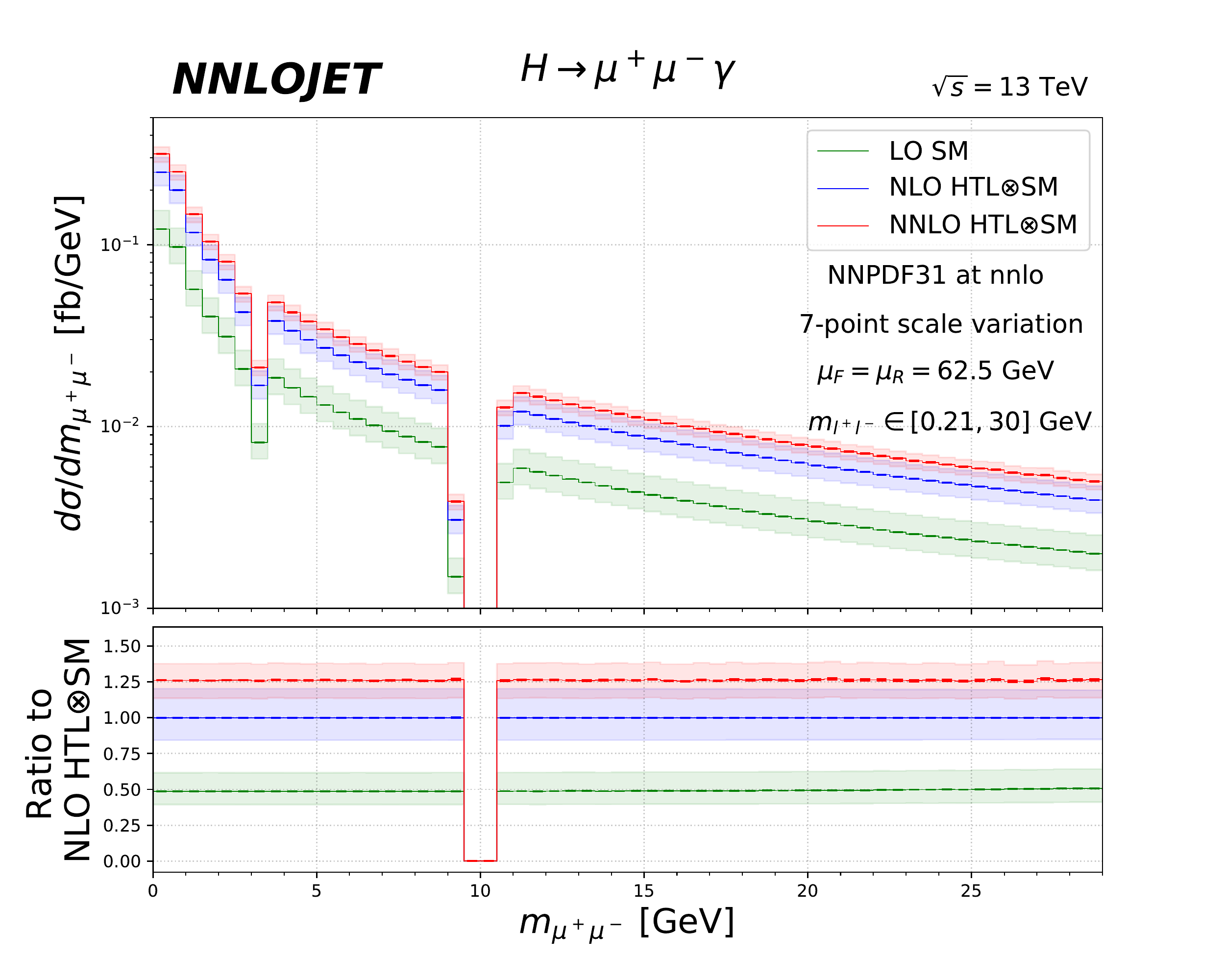}
    \end{subfigure}
    \\
    \begin{subfigure}[b]{0.49\textwidth}
        \includegraphics[width=\textwidth]{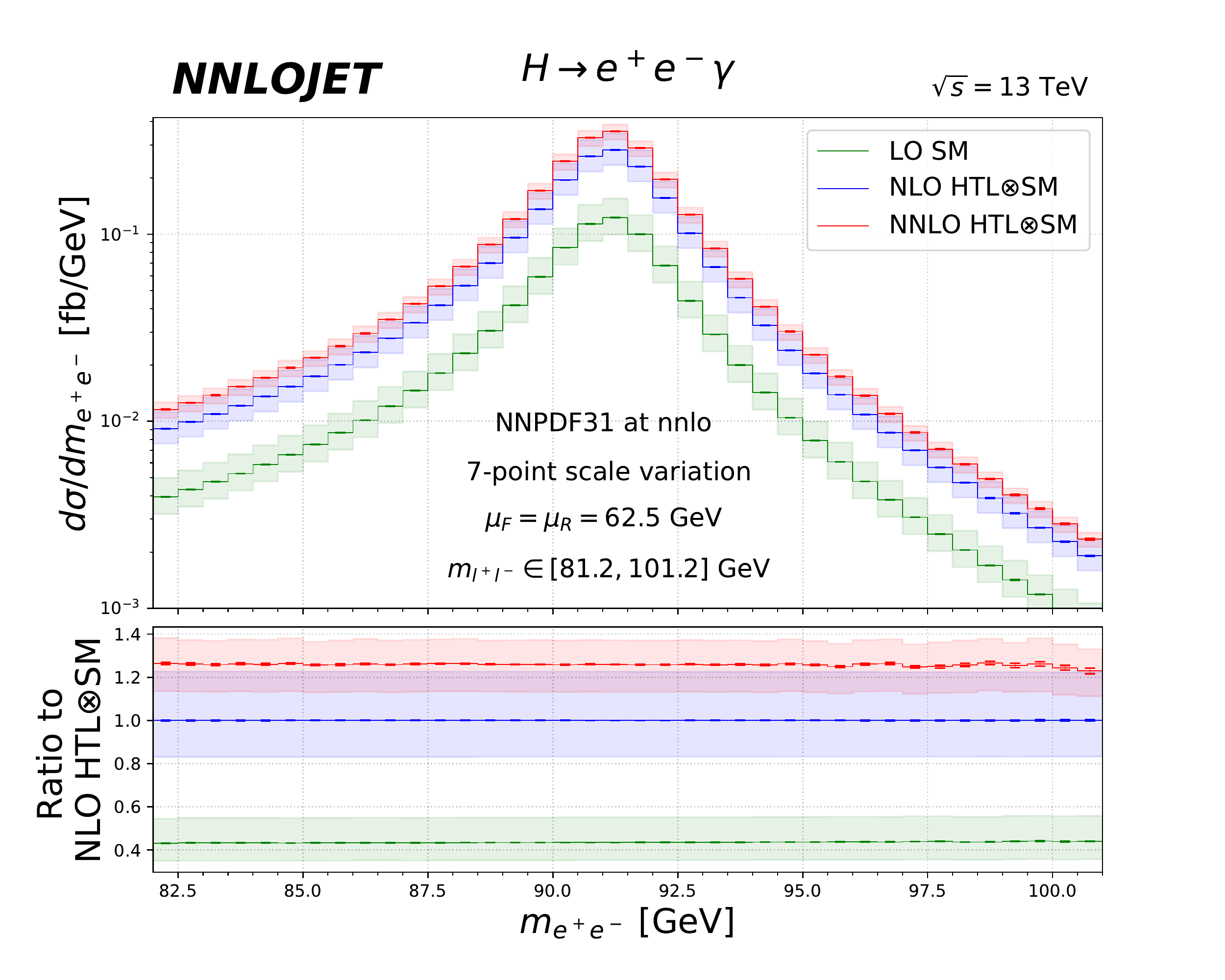}
    \end{subfigure}
    \begin{subfigure}[b]{0.49\textwidth}
        \includegraphics[width=\textwidth]{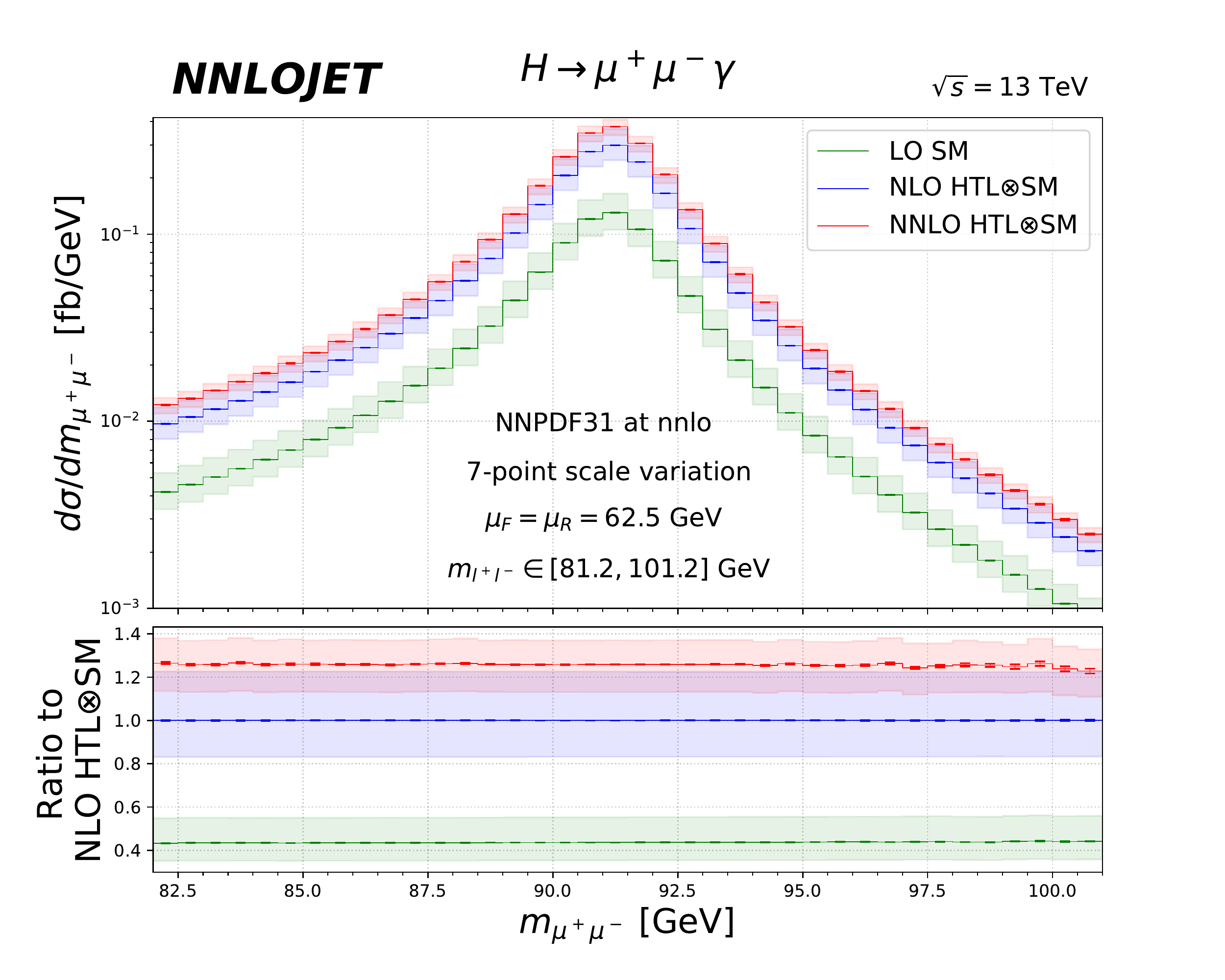}
    \end{subfigure}
    \caption{Fiducial invariant mass distribution of lepton pair
in low-mass (upper) and resonant-Z (lower) regions
      for electrons (left) and muons (right).}\label{fig:mll}
\end{figure}
Figure~\ref{fig:mll} displays the invariant mass distributions of the SFOS lepton pair. The top row presents the distribution in the low-mass region while the bottom row contains results for the resonant-Z region. The left and right columns are for the electron and muon final state leptons with different fiducial cuts listed in Table~\ref{tab:fid-cut}. The peak of each distribution is either near the on-shell photon or the Z boson pole. There are two gaps in the distribution for the low-mass region due to the mass veto cuts to remove the resonant contributions from $J/\psi$ and $\Upsilon$ decay products.
In the resonant-Z region, the  slope of the $m_{l^+l^-}$ distribution drops faster above the Z-pole than below.
With similar distance from the on-shell Z boson mass, the signal yield around 81.2~GeV is about six times larger than the signal yield at 101.2~GeV. This difference can be explained by the large $m_{l^+l^-}$ value competing with the final state photon phase space while $m_{l^+l^-\gamma}$  is constrained by the Higgs boson mass.
The lower panel of each subfigure displays the ratio of the fixed order contributions to the central value of the
NLO HTL$\otimes$SM prediction. It is found that the NNLO corrections
and their uncertainties are constant in $m_{l^+l^-}$, reproducing the values observed for the total fiducial cross sections given in
Table~\ref{tab:totalXS}.

\begin{figure}
    \centering
    \begin{subfigure}[b]{0.49\textwidth}
        \includegraphics[width=\textwidth]{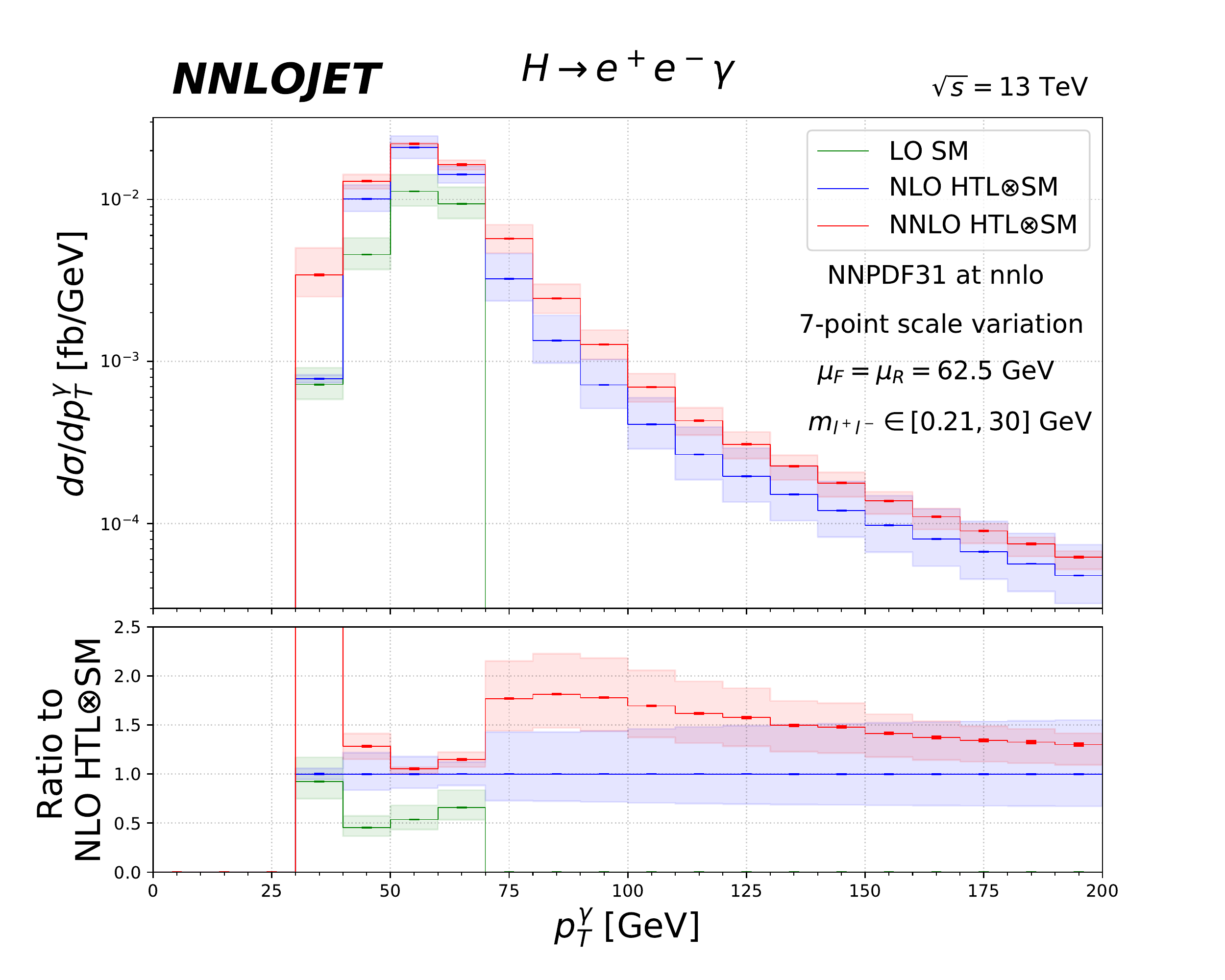}
    \end{subfigure}
    \begin{subfigure}[b]{0.49\textwidth}
        \includegraphics[width=\textwidth]{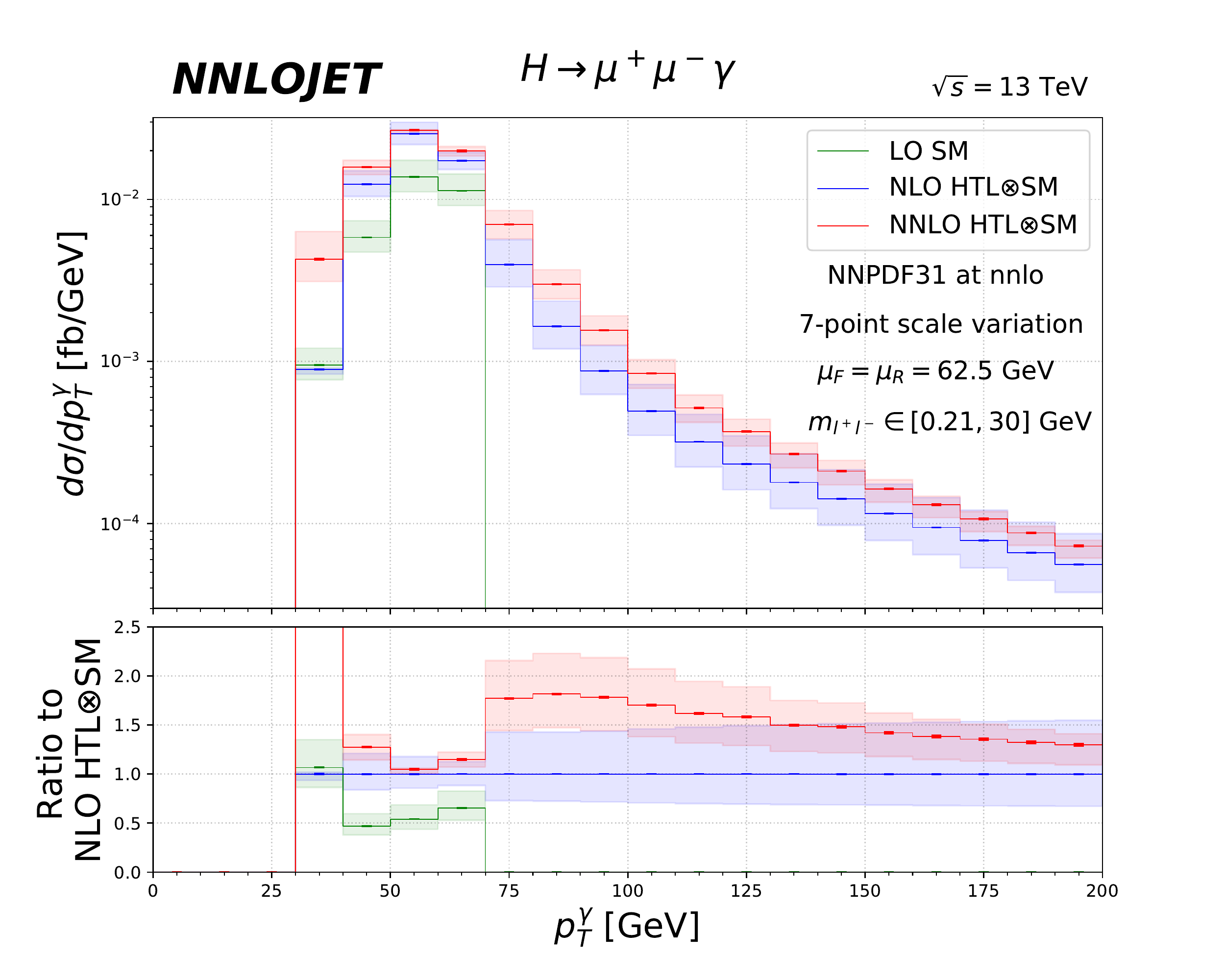}
    \end{subfigure}
    \\
    \begin{subfigure}[b]{0.49\textwidth}
        \includegraphics[width=\textwidth]{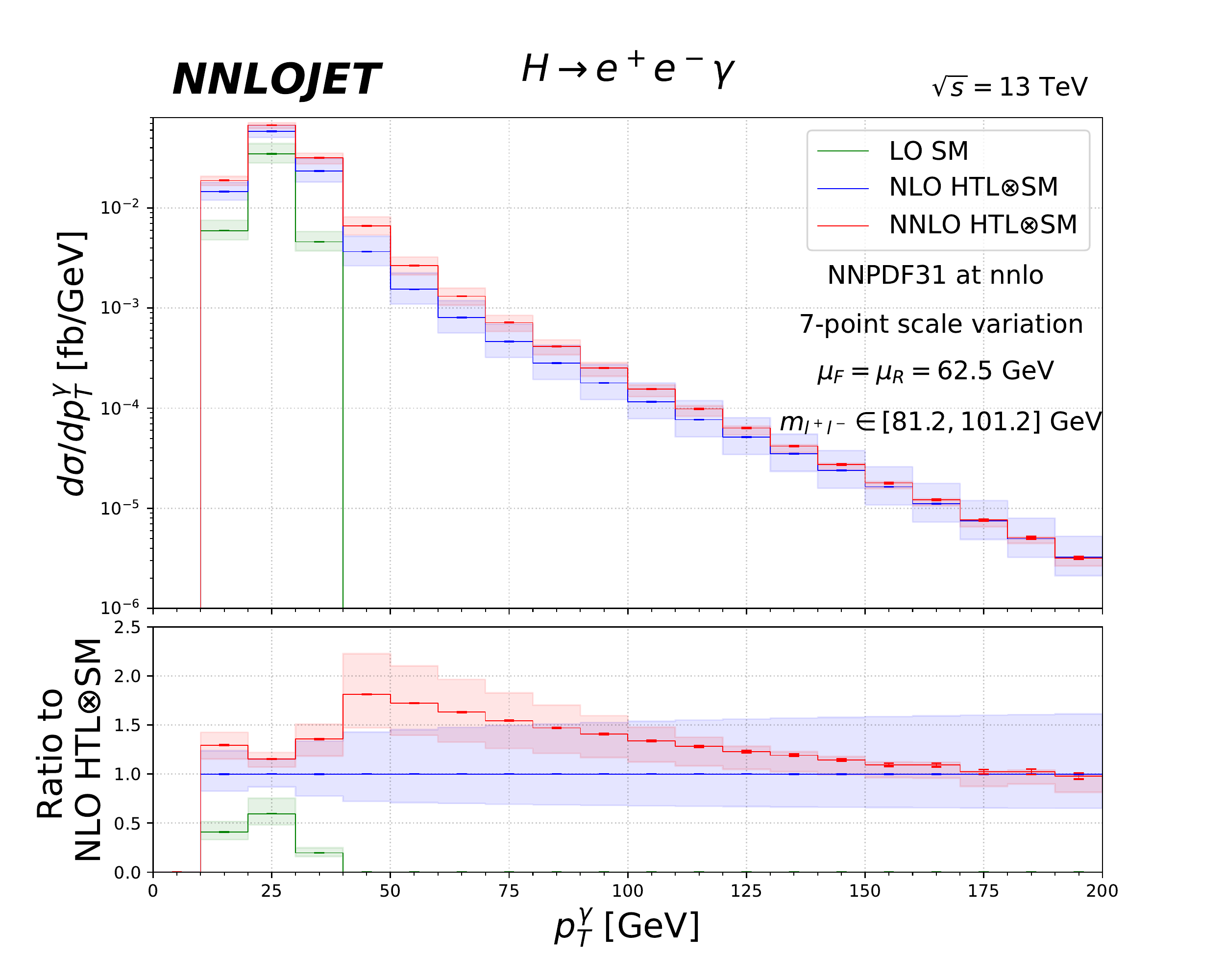}
    \end{subfigure}
    \begin{subfigure}[b]{0.49\textwidth}
        \includegraphics[width=\textwidth]{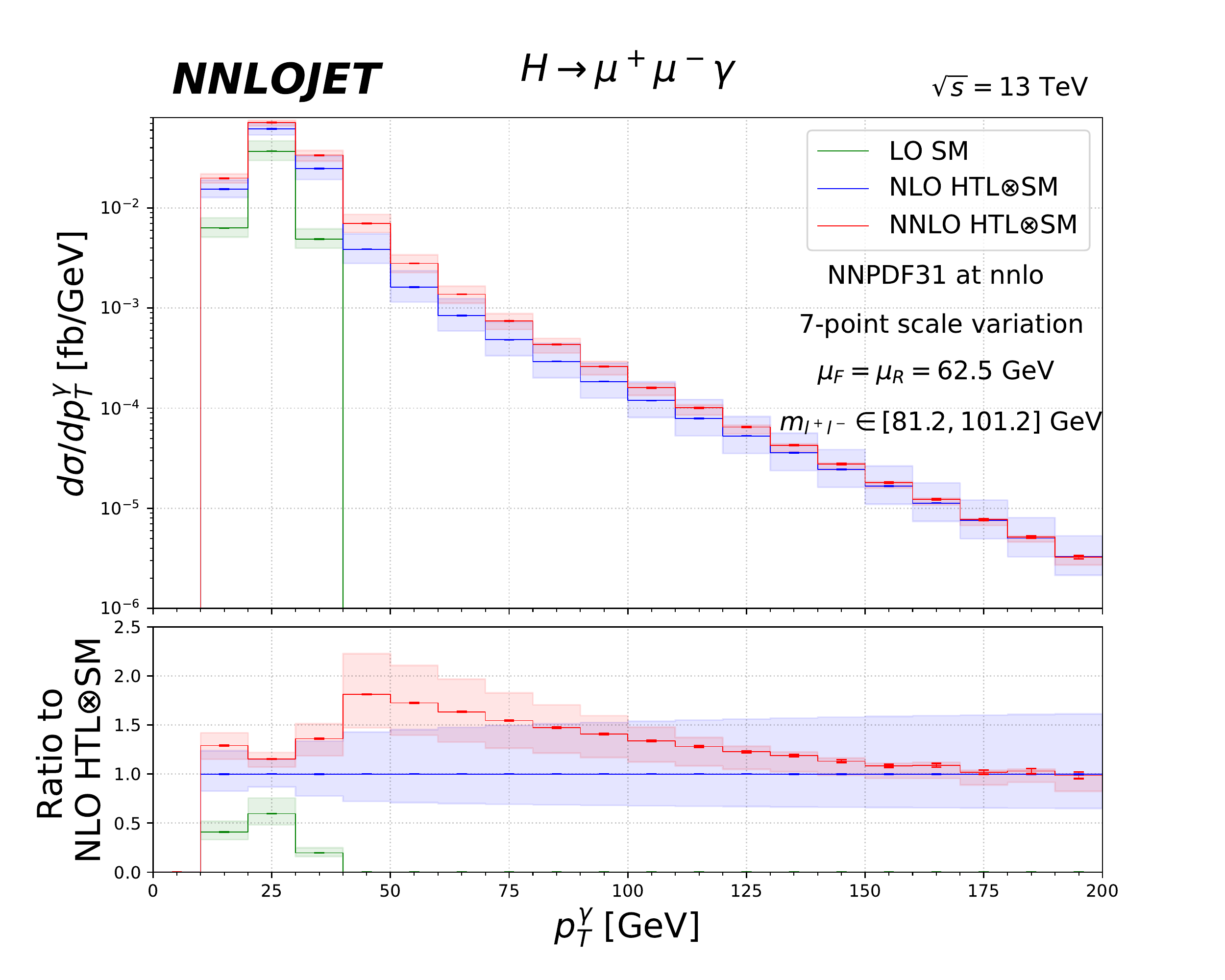}
    \end{subfigure}
    \caption{Fiducial transverse momentum distribution of the final state photon 
in low-mass (upper) and resonant-Z (lower) regions
      for electrons (left) and muons (right).}\label{fig:ptg}
\end{figure}

Transverse momentum distributions of the final state photon are summarized in Figure~\ref{fig:ptg}. By examining the LO results, we see that the distribution is constrained within the range between the minimum fiducial $p_T^\gamma$ cut and a
maximal value determined by the kinematical mass window.
The maximum $p_T^\gamma$ could reach $\sim65$~GeV in the low-mass range while
only extending up to $\sim35$~GeV within the kinematic constraints for the resonant-Z region.
This phase space constraint is lifted when including higher order QCD corrections due to initial state radiation,
which transfers a transverse momentum recoil to the $l^+l^-\gamma$ system. 
We observe a further strong enhancement in the distribution from NLO to NNLO by 80\%
just beyond the LO phase space edge due to the additional radiation allowed at NNLO, which is
accompanied by a large scale uncertainty even at NNLO. 
This Sudakov shoulder effect~\cite{Catani:1997xc} of large perturbative corrections at the edge of a Born-forbidden
phase space region is well-understood.
This enhancement attenuates with increasing $p_T^\gamma$, where also a reduced scale uncertainty is observed.
The scale variation band of the $p_T^\gamma$ distribution starts to completely overlap with NLO above 100~GeV for the resonant-Z
 region while the same behaviour only happens above 160~GeV  for the low-mass region.
The $p_T^\gamma$ distribution peaks at 55~GeV for the low-mass region while
only at 25~GeV for the resonant-Z region. This distinct difference allows the experimental analysis to apply higher
$p_T^\gamma$ cuts for the low-mass region with small penalty on the signal yield.
In the low-mass range, we do moreover observe very large NNLO corrections in the lowest $p_T^\gamma$ bin. These are a
result of the symmetric cuts on $p_T^\gamma$ and $p_T^{l^+l^-}$, which lead to implicit constraints on the direction
of soft real radiation at NLO, which are overcome at NNLO, resulting in the observed large corrections. Staged cuts
on  $p_T^\gamma$ and $p_T^{l^+l^-}$ can circumvent this problem, and should be applied in future experimental studies. 
\begin{figure}
    \centering
    \begin{subfigure}[b]{0.49\textwidth}
        \includegraphics[width=\textwidth]{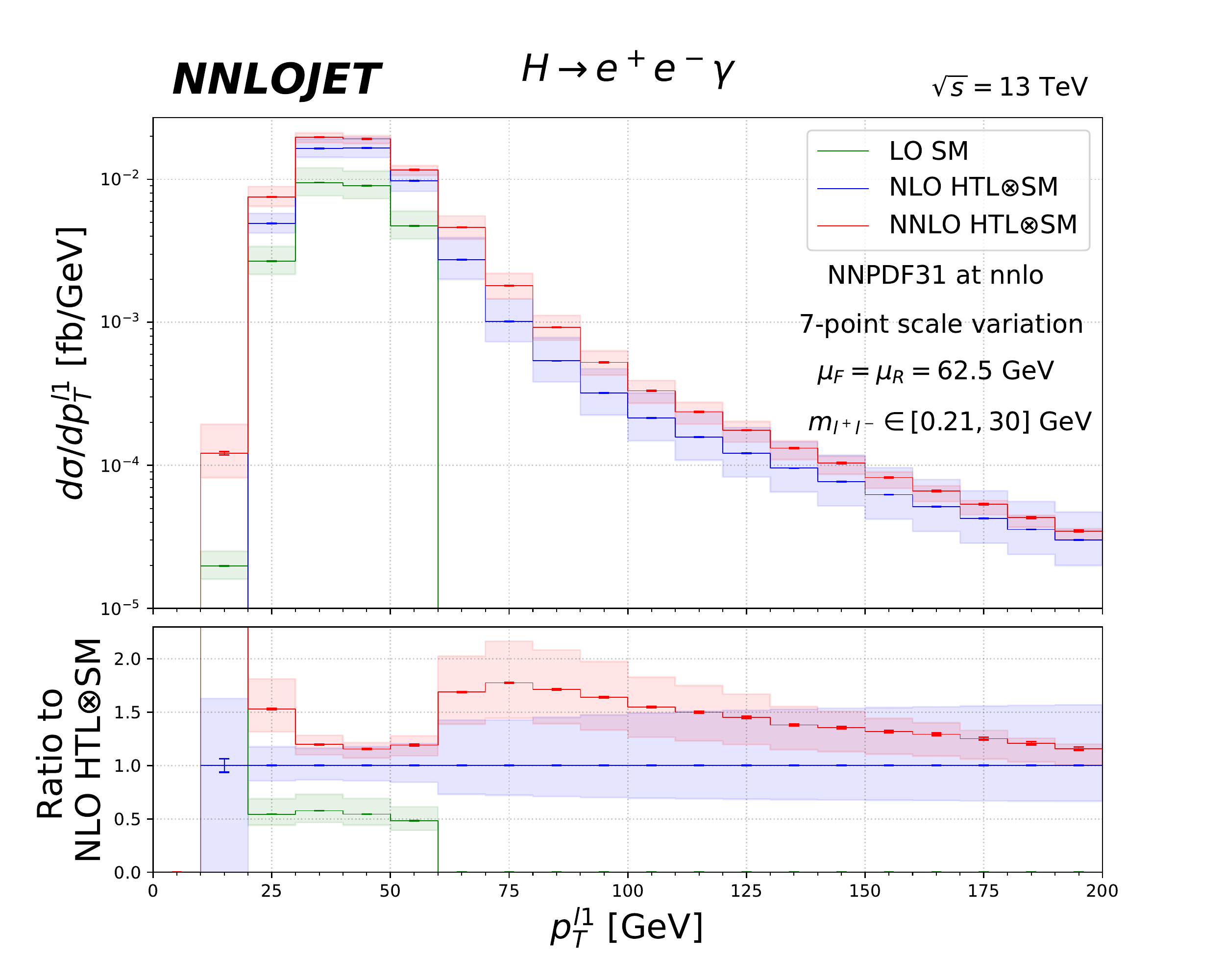}
    \end{subfigure}
    \begin{subfigure}[b]{0.49\textwidth}
        \includegraphics[width=\textwidth]{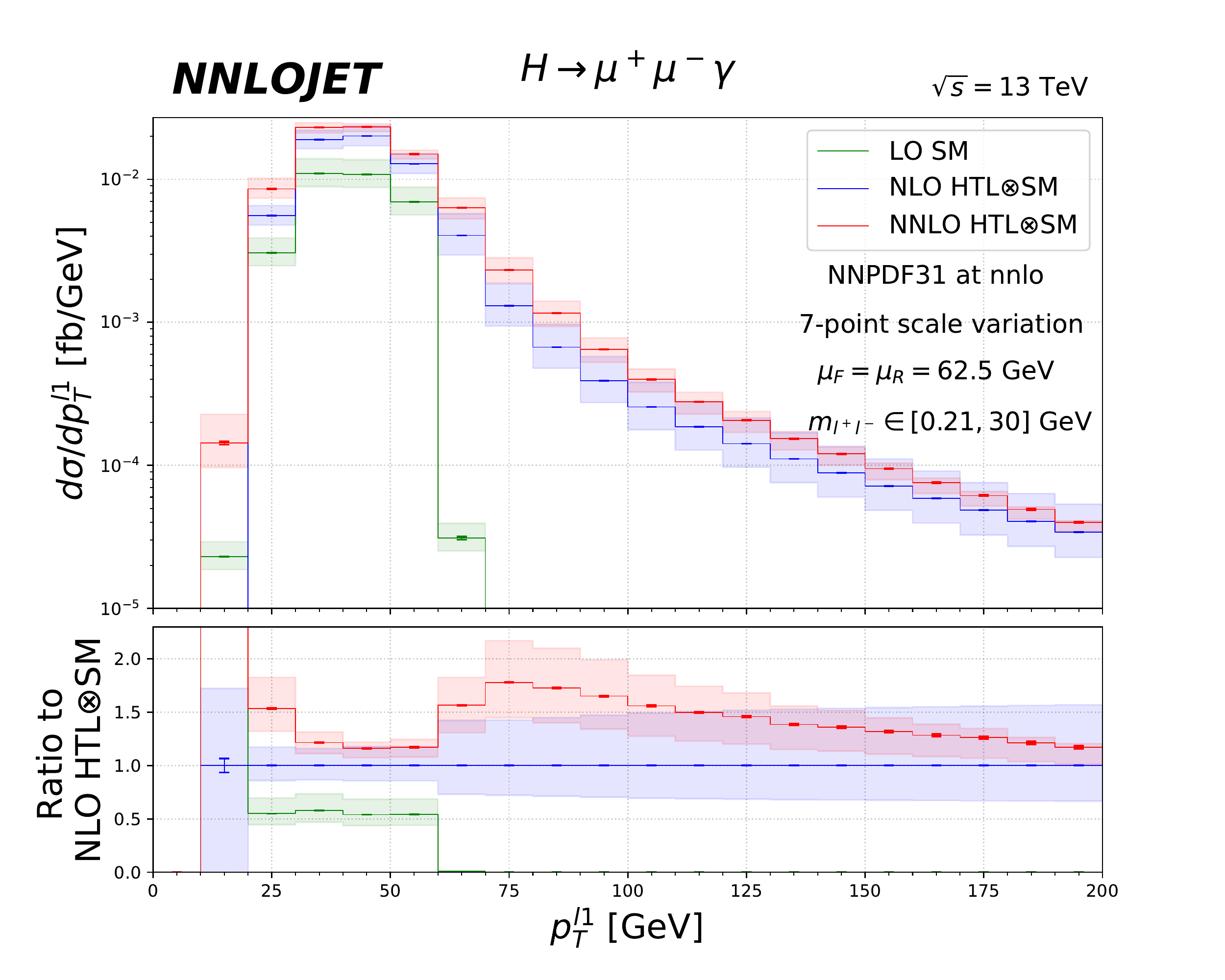}
    \end{subfigure}
    \\
    \begin{subfigure}[b]{0.49\textwidth}
        \includegraphics[width=\textwidth]{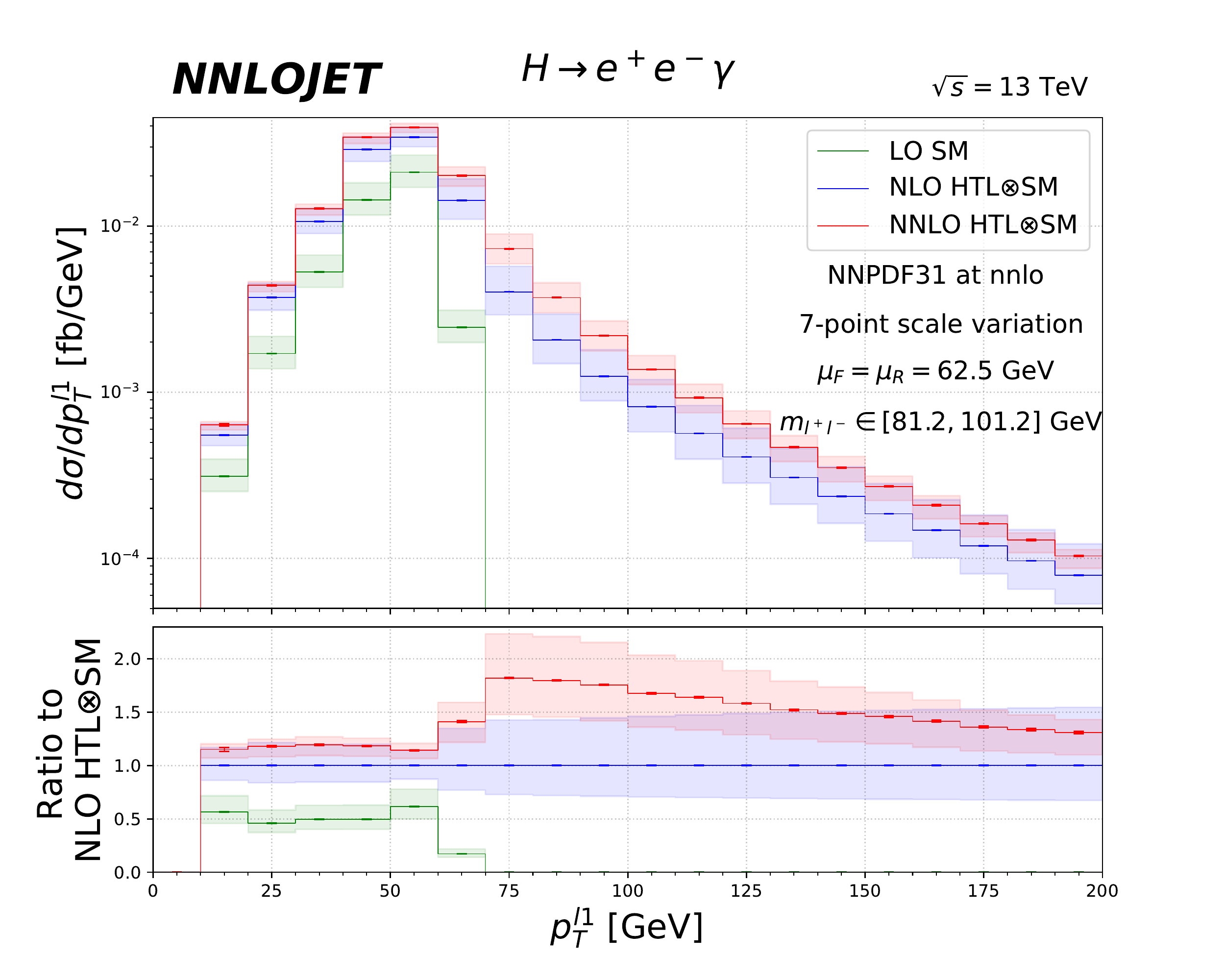}
    \end{subfigure}
    \begin{subfigure}[b]{0.49\textwidth}
        \includegraphics[width=\textwidth]{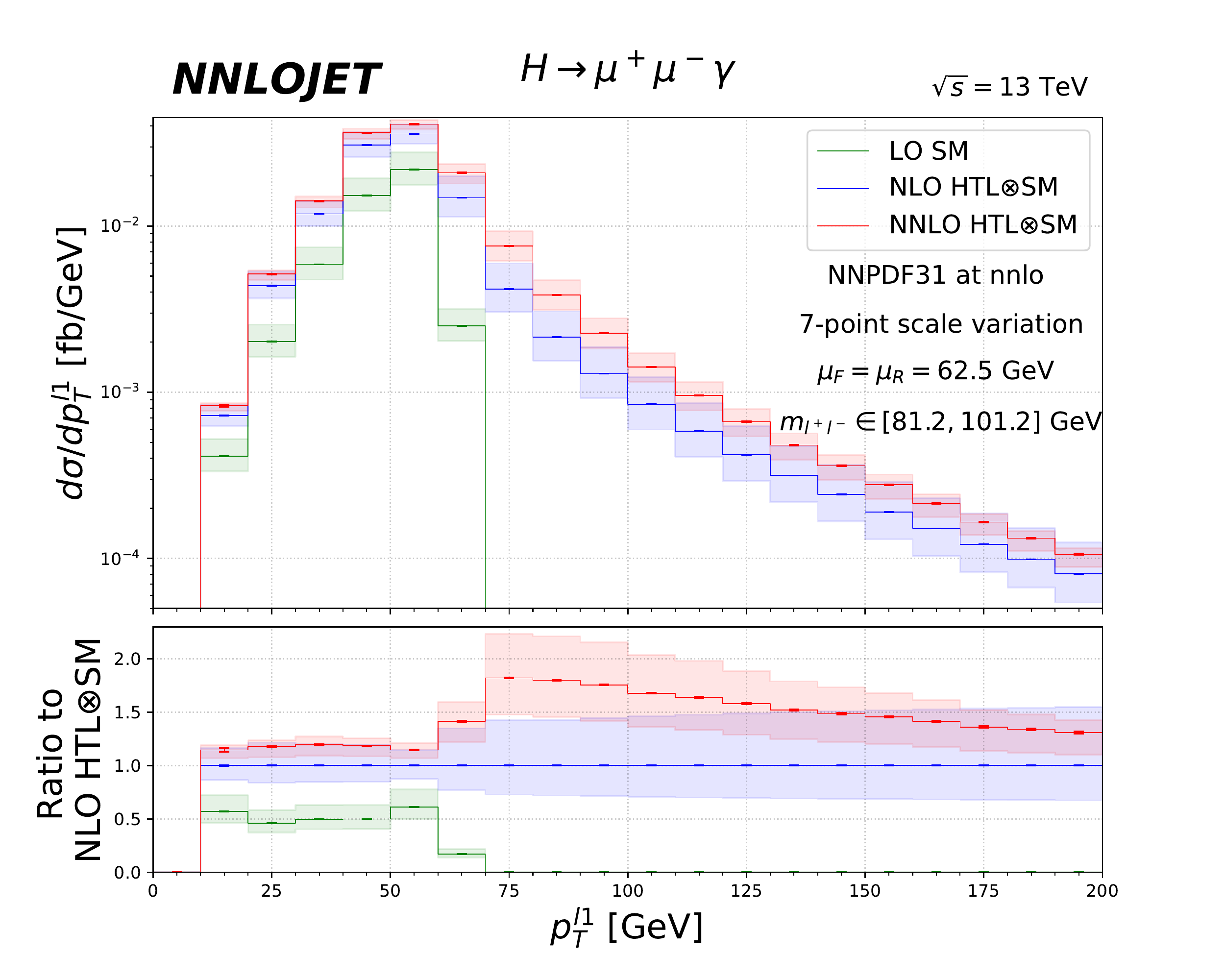}
    \end{subfigure}
    \caption{Transverse momentum distribution of the leading lepton in low-mass and Z boson resonant regions.}
    \label{fig:ptl1}
\end{figure}

The transverse momentum distribution of the leading lepton in Figure~\ref{fig:ptl1} shares many similarities with the $p_T^\gamma$ distributions in Figure~\ref{fig:ptg}. Phase space restrictions appear at LO and
constrain  $p_T^{l1}$ to be below 62.5~GeV in both mass regions (corresponding to a kinematical situation where the leading lepton recoils against the subleading lepton and the photon), and 
    the lepton--photon isolation as well as the low-mass $m_{e^+e^-}$ window can further lower this upper bound.
NLO and NNLO corrections extend the allowed kinematics to larger $p_T^{l1}$ regions with a non-trivial NNLO/NLO ratio especially directly above the Born-level kinematic threshold. In the low-mass region, we also observe very large corrections in the first bin of the $p_T^{l1}$ spectrum, again
resulting from the symmetric cuts on $p_T^\gamma$ and $p_T^{l^+l^-}$.
The peak of the $p_T^{l1}$ distribution is at 30~GeV for the low-mass region while being at 50~GeV for the resonant-Z region. The difference in fiducial cuts for the \ppheeg and \pphmumug channels results in an increase of about 20\% in the signal yield for the $\mu^+\mu^-$ channel in the low-mass region. The difference only appears as a normalization effect with negligible impact to the shape of distributions.

\begin{figure}
    \centering
    \begin{subfigure}[b]{0.49\textwidth}
        \includegraphics[width=\textwidth]{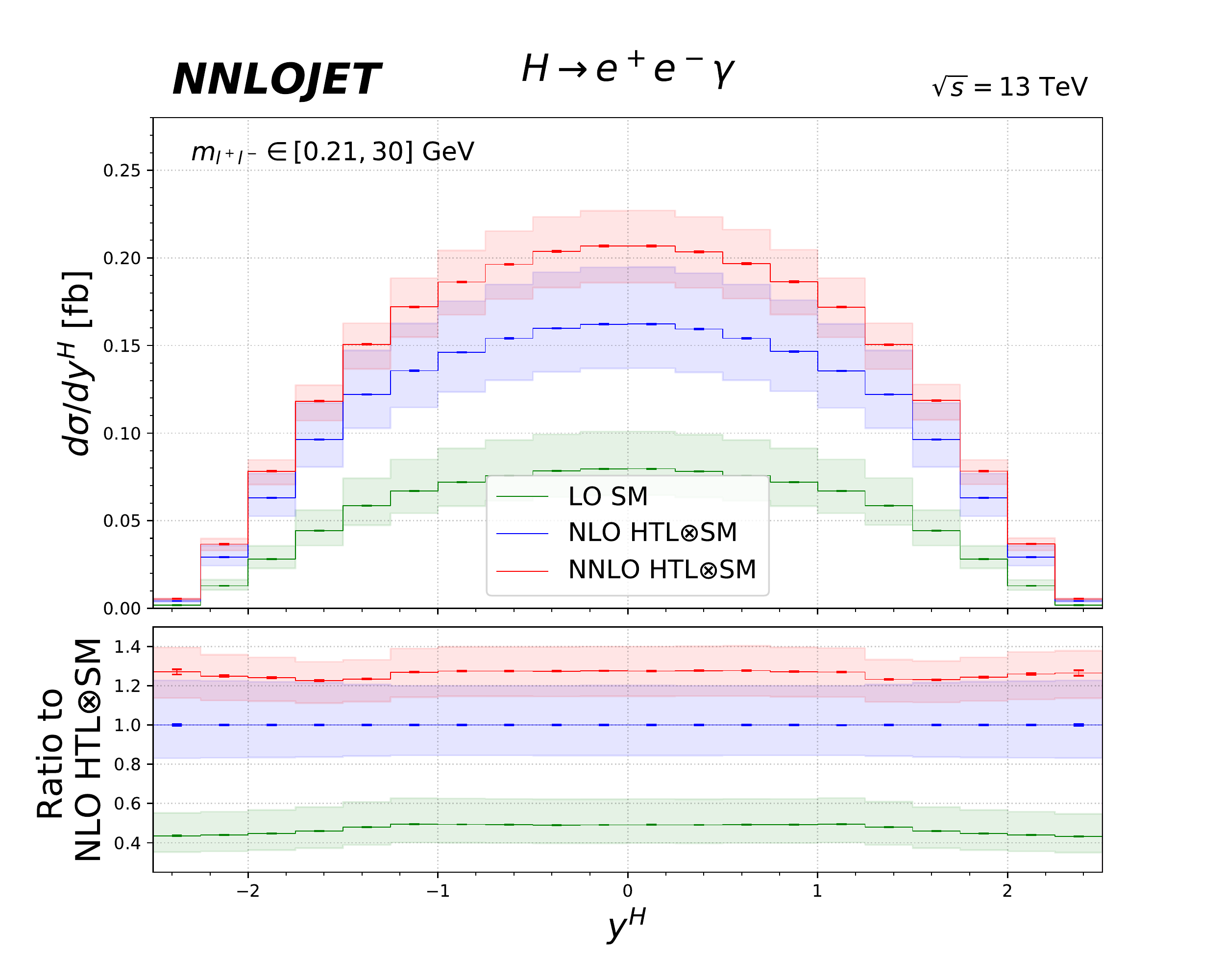}
    \end{subfigure}
    \begin{subfigure}[b]{0.49\textwidth}
        \includegraphics[width=\textwidth]{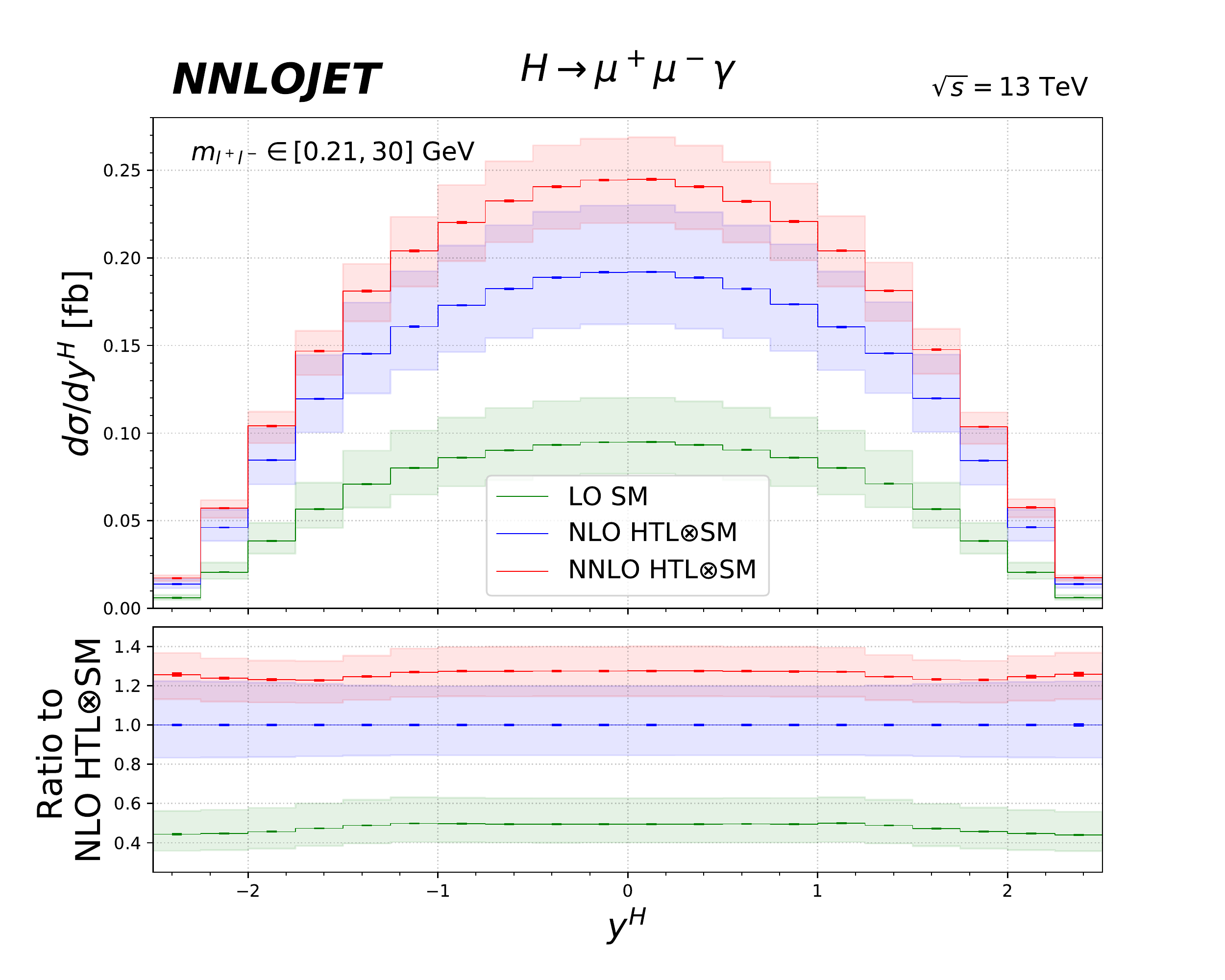}
    \end{subfigure}
    \\
    \begin{subfigure}[b]{0.49\textwidth}
        \includegraphics[width=\textwidth]{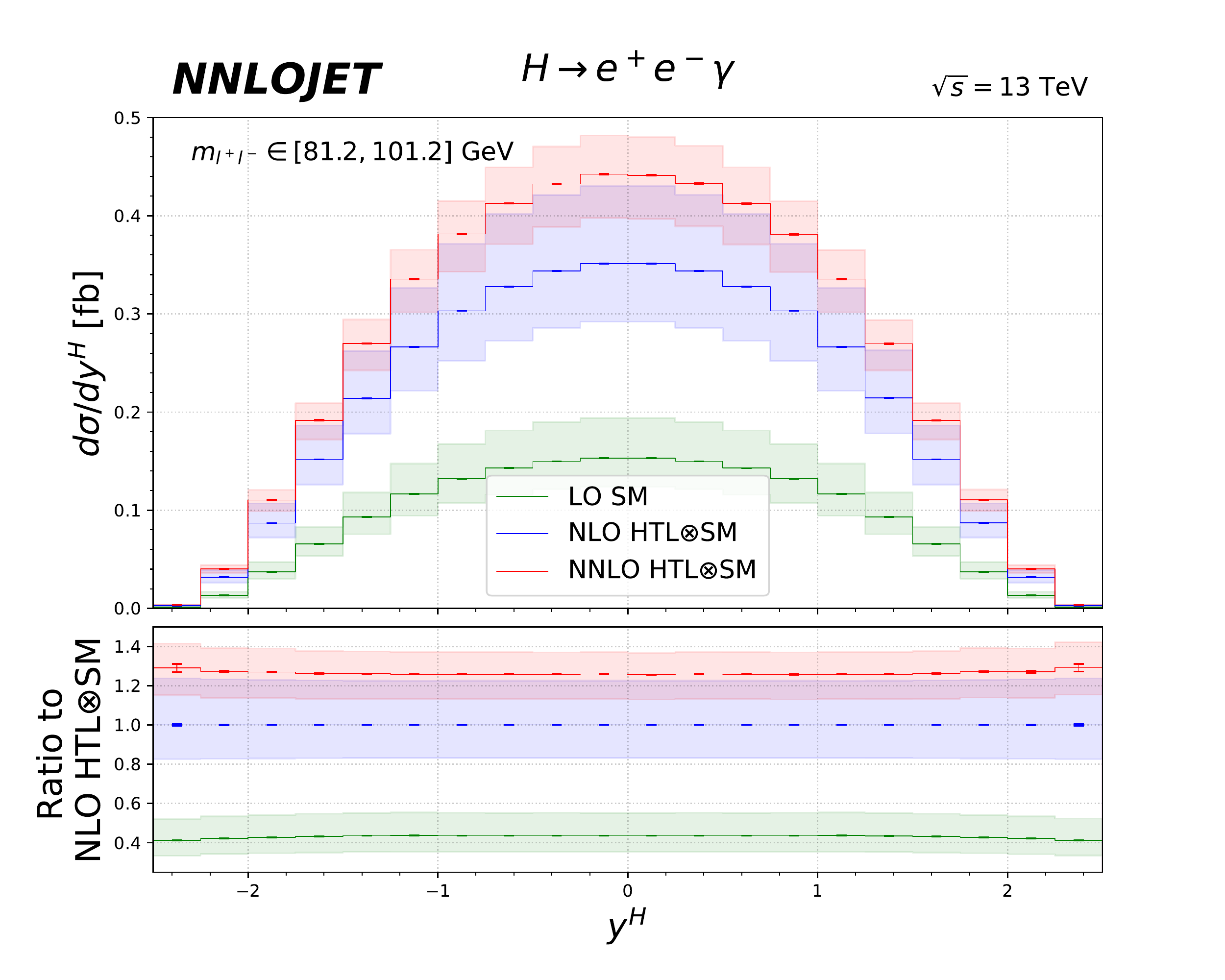}
    \end{subfigure}
    \begin{subfigure}[b]{0.49\textwidth}
        \includegraphics[width=\textwidth]{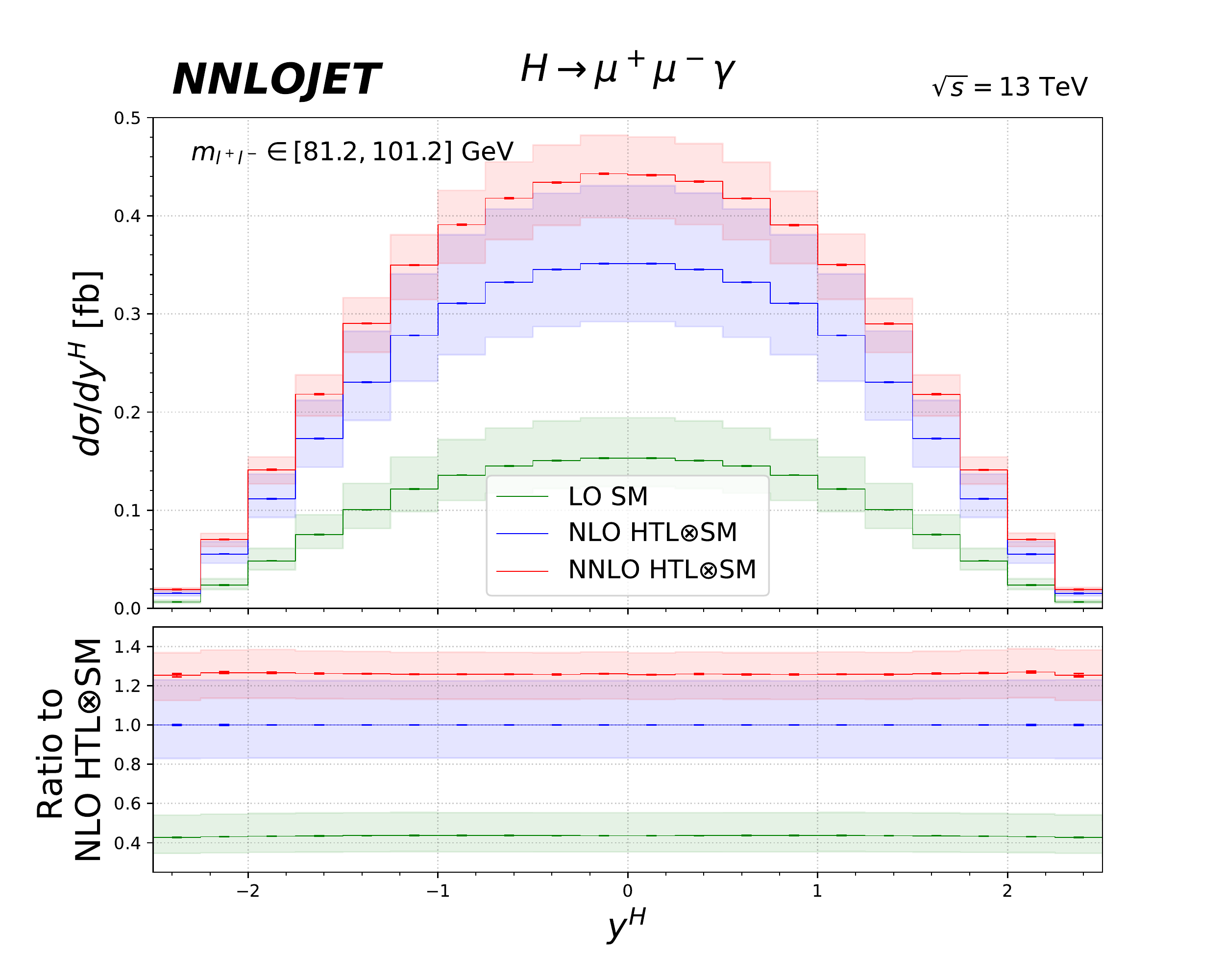}
    \end{subfigure}
    \caption{Fiducial rapidity distribution of the Higgs boson
in low-mass (upper) and resonant-Z (lower) regions
      for electrons (left) and muons (right).}\label{fig:yh}
\end{figure}
Figure~\ref{fig:yh} displays the Higgs boson rapidity distributions. We observe flat differential K-factors at NLO and NNLO
in the resonant-Z region. 
The scale variation bands partially overlap between NLO and NNLO predictions with +25\% correction at NNLO.
For the low-mass region, there is a small modulation of radiative corrections around $y^H\sim\pm1.3$ due to
kinematical effects. For di-lepton production near the on-shell photon pole, the \pphllg decay channel is effectively
a two-body decay with massless final states. Again, the symmetric fiducial cuts on $p_T^\gamma$ and $p_T^{l^+l^-}$
introduce implicit constraints on the Born-level kinematics. In the centre-of-mass (COM) frame
of Higgs boson production, the rapidity of final state photon is $|y^\gamma|_{\text{COM}} = \cosh^{-1}(m_H/(2p_T^\gamma))$ which leads to $|y^\gamma|_{\text{COM}} \lesssim 1.1$ for $p_T^\gamma>37.5$~GeV. This constraint is overcome
  by higher-order QCD radiation. 
  Due to the non-vanishing longitudinal momemtum of the Higgs boson, this does not translate into a sharp edge in the laboratory frame, it does nevertheless induce a variation of the higher order corrections with the Higgs rapidity at approximately $|y^{H}| \lesssim |y^\gamma|_\text{max} - |y^\gamma|_{\text{COM}} \sim 1.3$.
  A similar effect has been discussed recently in~\cite{Ebert:2019zkb,Chen:2021isd,Salam:2021tbm} for the $H\rightarrow\gamma\gamma$ decay channel. Away from the on-shell photon pole, the genuine three-body decay kinematics protect against this
  effect, which fully disappears from the $y^H$ distributions in the resonant-Z region.

\begin{figure}
    \centering
    \begin{subfigure}[b]{0.49\textwidth}
        \includegraphics[width=\textwidth]{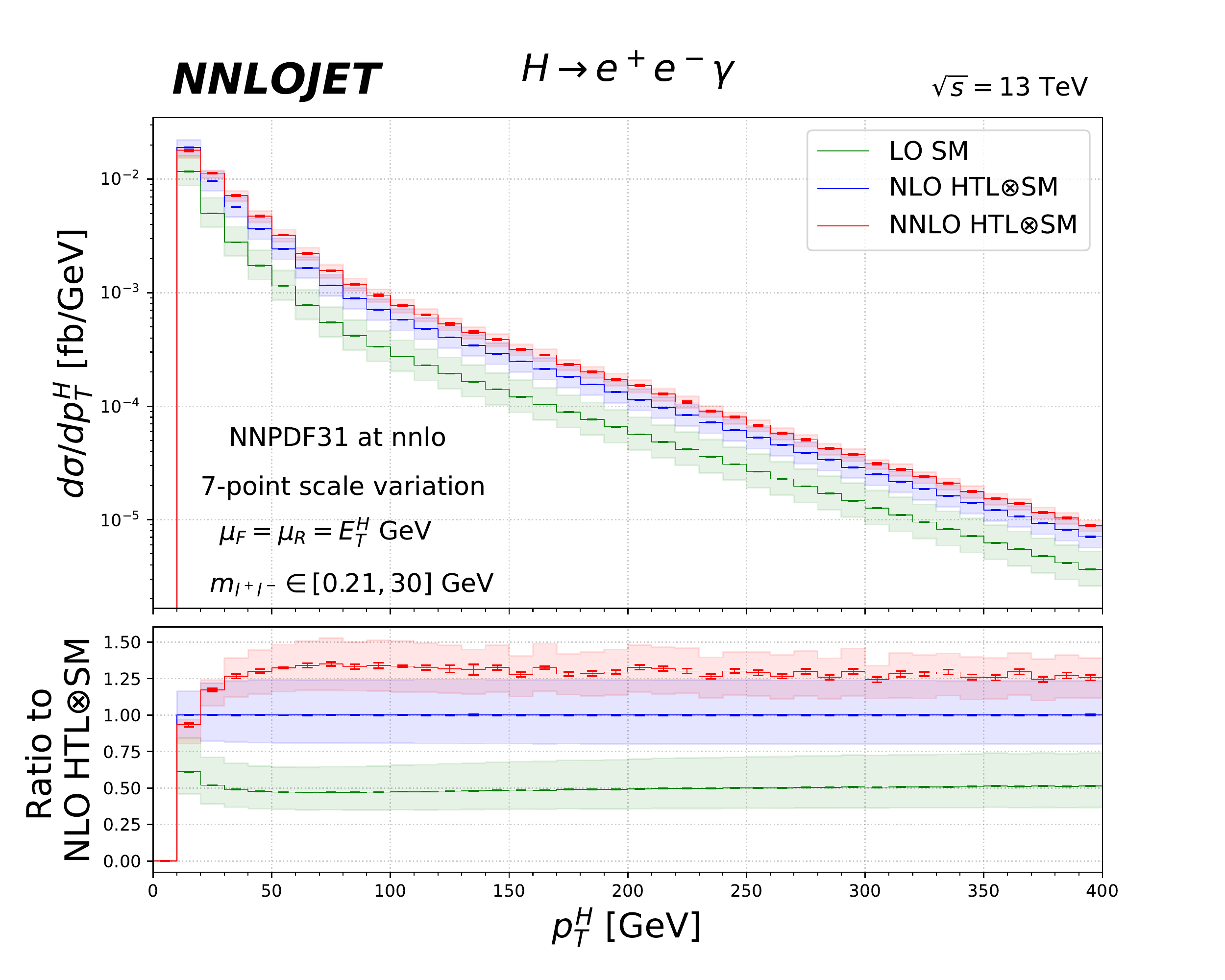}
    \end{subfigure}
    \begin{subfigure}[b]{0.49\textwidth}
        \includegraphics[width=\textwidth]{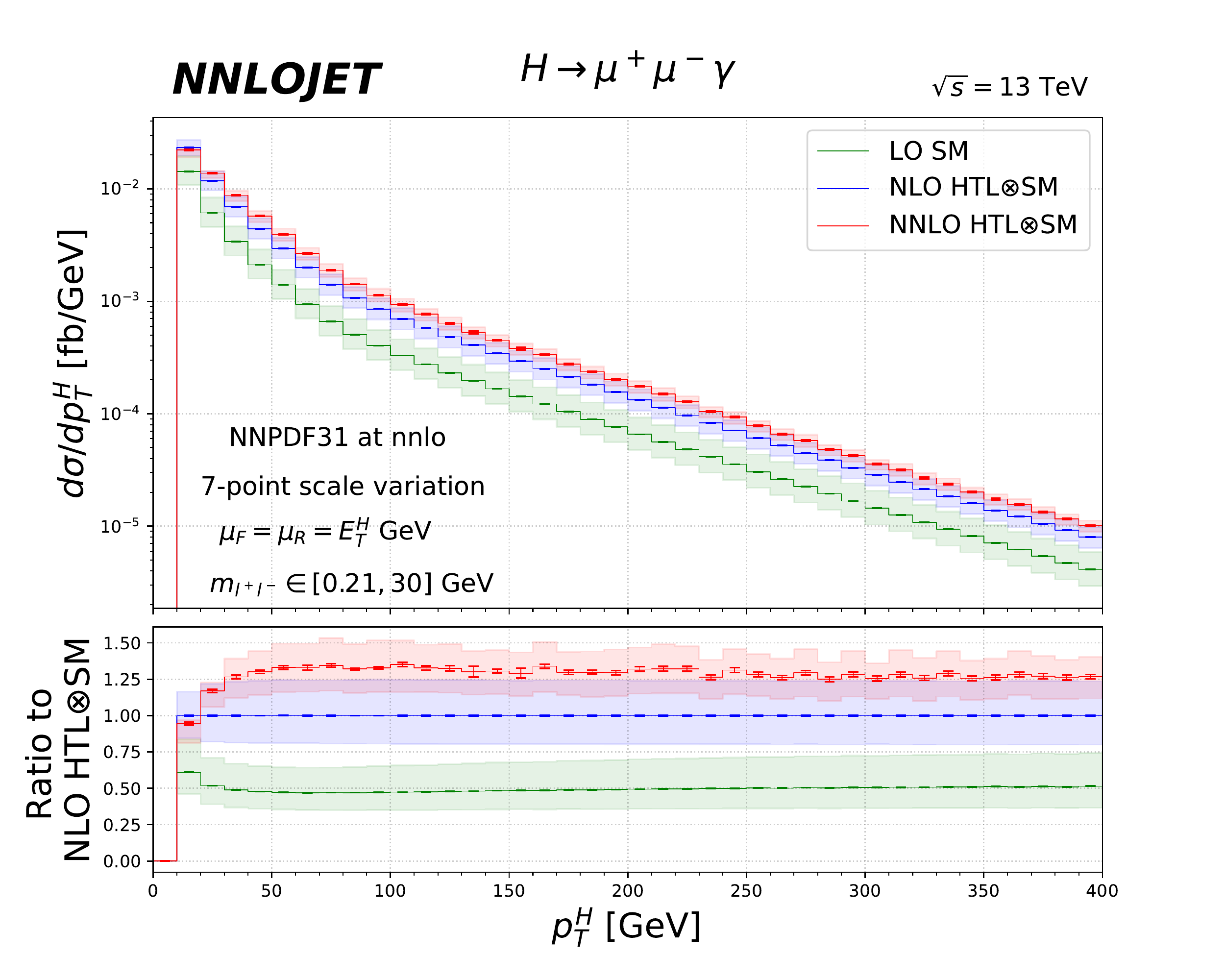}
    \end{subfigure}
    \\
    \begin{subfigure}[b]{0.49\textwidth}
        \includegraphics[width=\textwidth]{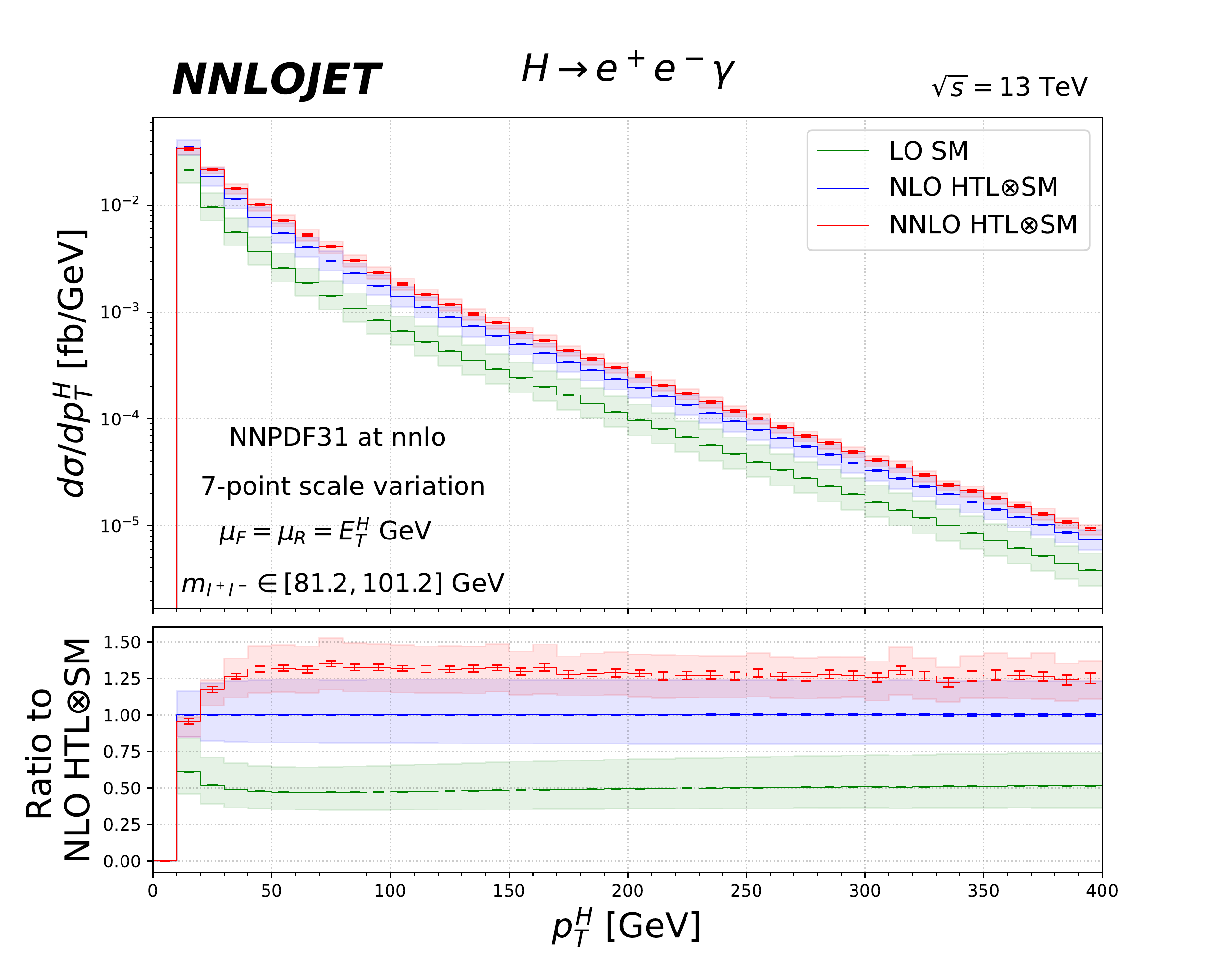}
    \end{subfigure}
    \begin{subfigure}[b]{0.49\textwidth}
        \includegraphics[width=\textwidth]{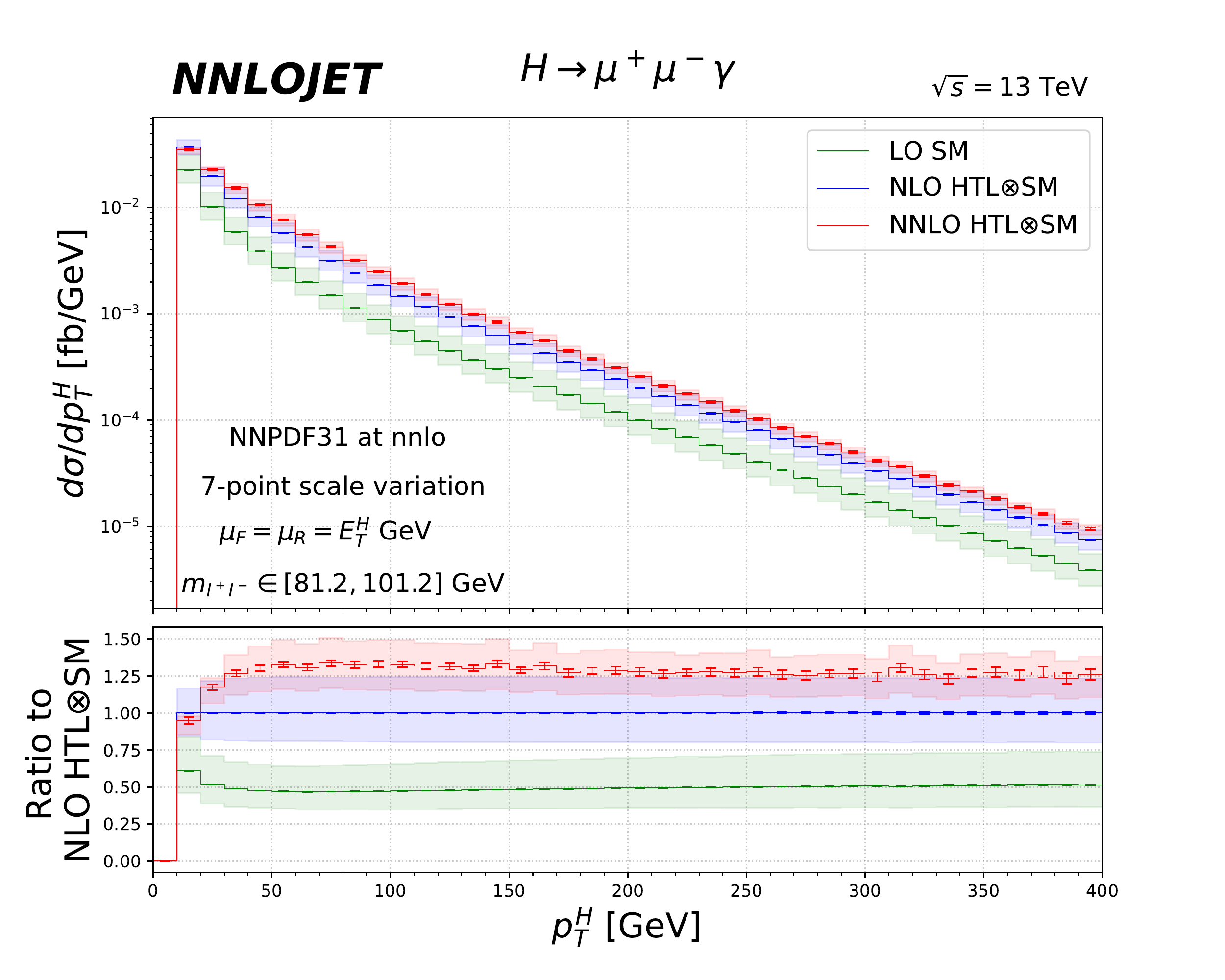}
    \end{subfigure}
    \caption{Fiducial transverse momentum distribution of Higgs boson in low-mass (upper) and resonant-Z (lower) regions
      for electrons (left) and muons (right).}\label{fig:pth}
\end{figure}
The predictions for the  Higgs transverse momentum distributions in the  \pphllg channel are shown in Figure~\ref{fig:pth}.
The Born-level kinematics of inclusive Higgs boson production correspond to vanishing transverse momentum.
Consequently, the fixed-order prediction of the transverse momentum distribution requires one extra power of
$\alpha_s$, as compared to inclusive production. Our calculation of the $p_T^H$
distribution is thus is based on Higgs-plus-jet production, with the jet requirement replaced by a minimum cut on
$p_T^H> 10$~GeV.
A consistent counting of orders is applied, with the LO $p_T^H$ distribution corresponding to the NLO contribution
to inclusive production in the previous figures. 
The NNLO QCD correction is based on our previous work~\cite{Chen:2016zka,Chen:2018pzu,Bizon:2018foh}, extended to the \pphllg decay channels. We use a dynamical central scale choice of $E_T^H=\sqrt{m_H^2+(p_T^H)^2}$ in the calculation
of the $p_T^H$ distribution to improve the perturbative convergence in the
large transverse momentum region.
For both low-mass and resonant-Z region, we observe flat differential K-factors from higher order corrections above
$p_T^H=$50~GeV. There is a 30--35\% enhancement from NLO to NNLO with scale uncertainties reducing from about $\pm 22\%$
to $\pm13\%$. For the $p_T^H$ distribution below 30~GeV, transverse momentum resummation effects up to N$^3$LL should be
considered~\cite{Chen:2018pzu,Bizon:2018foh,Billis:2021ecs}, which are beyond the scope of this study. 

The differences that were observed in the shapes of the $m_{l^+l^-}$, $p_T^\gamma$ and $p_T^{l1}$ distributions between electrons and muons are not present in the $p_T^H$ and $y^H$ distributions for either low-mass and resonant-Z regions. This will be
an advantage when combining various Higgs decay channels to reconstruct inclusive Higgs observables. 

\section{Conclusions and outlook}
\label{sec:conc}

The rare Higgs boson decay mode \pphllg can provide important information on the Higgs boson couplings and is particularly sensitive 
to physics effects beyond the Standard Model. Searches for this decay mode~\cite{ATLAS:2020qcv,ATLAS:2021wwb} 
concentrate on two lepton invariant mass ranges (low-mass 
and resonant-Z), with first evidence reported recently~\cite{ATLAS:2020qcv} in the low-mass range.

In this paper, we investigated the impact of QCD corrections up to NNLO on fiducial cross sections related to the \pphllg decay mode for final-state 
electrons and muons, 
by combining this decay mode with the NNLO calculations for Higgs and Higgs-plus-jet final states. 
Higher order QCD corrections were found to be quite uniform for the distributions in the lepton pair invariant mass and in the 
Higgs boson transverse momentum  and rapidity.  Non-trivial kinematical features are observed in QCD corrections to the transverse momentum
distributions of the leading lepton and the photon,  which can be understood from a non-trivial interplay between the 
fiducial event selection cuts and higher-order QCD radiation. These effects are particularly pronounced in the low-mass region, 
where they lead to sizeable QCD corrections to the acceptance factors for electrons and muons,
which decrease at higher orders while remaining constant in 
the resonant-Z region. Several perturbative instabilities (Sudakov shoulders) are observed in these distributions, which could 
be partly eased by the introduction of asymmetric cuts on the final state particles.

Our results can contribute to the optimisation of searches for the  \pphllg decay mode and will subsequently enable its precision study
on the increasing LHC data set.

\acknowledgments

We would like to thank Sarah Heim and Anthony Morley for interesting discussions on the ATLAS studies~\cite{ATLAS:2020qcv,ATLAS:2021wwb} 
of the \pphllg decay mode.
The authors also want to thank Juan Cruz-Martinez, James Currie, Aude Gehrmann-De Ridder, Marius H\"ofer, Matteo Marcoli, Jonathan Mo, Tom Morgan, Joao Pires, Robin Sch\"urmann, Duncan Walker and James Whitehead for useful discussions and their many contributions to the \nnlojet code.
We thank the University of Zurich S3IT (http://www.s3it.uzh.ch) and Swiss National Supercomputing Centre (CSCS) with project ID UZH10 for providing support and computational resources.
This project has received funding from the European Research Council (ERC) under the European Union's Horizon 2020 research and innovation programme grant agreement 101019620 (ERC Advanced Grant TOPUP), from the UK Science and Technology Facilities Council (STFC) through grant ST/T001011/1, from the Swiss National Science Foundation (SNF) under contract 200020-175595, and from the Deutsche Forschungsgemeinschaft (DFG, German Research Foundation) under grant 396021762-TRR 257.

\end{document}